%% file: main_sn-article.tex
\documentclass[pdflatex,sn-mathphys]{sn-jnl}

\jyear{2023}%

\theoremstyle{thmstyleone}%
\theoremstyle{thmstyletwo}%

\theoremstyle{thmstylethree}%

\usepackage{tabu}                      
\usepackage{booktabs}                  

\newcommand{\jun}[1]{{\color{black} {#1}}}

\begin{document}

\title[TRIVEA]{\textbf{TRIVEA}: \textbf{T}ransparent \textbf{R}anking \textbf{I}nterpretation using \textbf{V}isual \textbf{E}xplanation of Black-Box \textbf{A}lgorithmic Rankers}


\author[1]{\fnm{Jun} \sur{Yuan}}\email{jy448@njit.edu}

\author[1]{\fnm{Kaustav} \sur{Bhattacharjee}}\email{kb526@njit.edu}

\author[2]{\fnm{Akm Zahirul} \sur{Islam}}\email{akm.islam@njit.edu}

\author*[2]{\fnm{Aritra} \sur{Dasgupta}}\email{aritra.dasgupta@njit.edu}

\affil[1]{\orgdiv{Department of Informatics}, \orgname{New Jersey Institute of Technology}, \orgaddress{
\country{USA}
}}

\affil*[2]{\orgdiv{Department of Data Science}, \orgname{New Jersey Institute of Technology}, \orgaddress{
\country{USA}
}}



\abstract{
\input{sections/abstract.tex}
}

\keywords{Visual Analytics, Learning-to-Rank, Explainable ML, Ranking}



\maketitle
\input{sections/introduction2}
    
\input{sections/relatedWork}

\input{sections/problemTask}

\input{sections/design}

\input{sections/results}

\input{sections/caseStudy}

\input{sections/subjectivefeedback}

\backmatter






\section*{Declarations}
\textbf{Conflict of interest/Competing interests}. No conflict of interest or competing interests.

\noindent \textbf {Data availability.} All data analysed during this study are included in this published article and its supplementary information files.








\begin{appendices}

\section{Section title of first appendix}\label{secA1}
Not Applicable




\end{appendices}


\bibliography{bib}


\end{document}

%% file: sections/abstract.tex
Ranking schemes drive many real-world decisions, like, where to study, whom to hire, what to buy, etc. Many of these decisions often come with high consequences. For example, a university can be deemed less prestigious if not featured in a top-k list, and consumers might not even explore products that do not get recommended to buyers. At the heart of most of these decisions are opaque ranking schemes, which dictate the ordering of data entities, but their internal logic is inaccessible or proprietary. Drawing inferences about the ranking differences is like a guessing game to the stakeholders, like, the rankees~(i.e., the entities who are ranked, like product companies) and the decision-makers~(i.e., who use the rankings, like buyers). 
In this paper, we aim to enable transparency in ranking interpretation by using algorithmic rankers that learn from available data and by enabling human reasoning about the learned ranking differences using explainable AI~(XAI) methods. To realize this aim, we leverage the exploration-explanation paradigm of human-data interaction to let human stakeholders explore subsets and groupings of complex multi-attribute ranking data using visual explanations of model fit and attribute influence on rankings. We realize this explanation paradigm for transparent ranking interpretation in \textit{TRIVEA}, a visual analytic system that is fueled by:
i) visualizations of model fit derived from algorithmic rankers that learn the associations between attributes and rankings from available data and
ii) visual explanations derived from XAI methods that help abstract important patterns, like, the relative influence of attributes in different ranking ranges.
Using TRIVEA, end users not trained in data science have the agency to transparently reason about the global and local behavior of the rankings without the need to open black-box ranking models and develop confidence in the resulting attribute-based inferences. We demonstrate the efficacy of TRIVEA using multiple usage scenarios and subjective feedback from researchers with diverse domain expertise.



%% file: sections/introduction2.tex
\section{Introduction}


Rankings are convenient heuristics for the human mind to make real-world choices. What we eat, shop, watch, study, etc. – rank-ordered lists of data entities, like restaurants, products, and universities, ubiquitously guide those decisions. 
However, many of these ranking schemes are often proprietary and inaccessible, yet, they have high consequences. For example, a university that is not on the top-$k$ list can be deemed as less prestigious; a product that is not recommended to buyers can lose substantial amounts in revenue; a job candidate who does not feature among the top applicants would not objectively know how to improve their chances relative to an applicant pool. From the perspective of stakeholders, like, data subjects who are ranked~(henceforth, termed as \textit{rankees}) or \textit{decision-makers}, it is often a guessing game for them to interpret the logic behind the ranking information that matters to them. 

Such inaccessibility and lack of transparency are ultimately detrimental to creating equitable socio-technical systems~\cite{bauer2009designing} where proprietary ranking schemes could be questionable yet, hold disproportionate power over stakeholders. Our work addresses this problem by conceptualizing an analytical workflow~(Figure~\ref{figs:overview}) that combines machine-learning explanations with expressive visualizations for making ranking schemes interpretable and actionable to different stakeholders. 
We learn a model by using the approach of supervised learning: training learning-to-rank~(LTR) algorithms on publicly available ranking data.  Then we use explanations of learned rankings to express associations between rank positions and attribute values. The learned rankings derived through modeling thus serve as the means to an end of discovering the signals in high-dimensional data spaces. These signals, capturing the attribute influence of rankings, need to be communicated effectively to end users. 
As opposed to score-based multi-attribute rankings, the challenge here is to express the learned scoring function faithfully and clearly. A concise mathematical formula may fail to capture and communicate variance in local data neighborhoods. We address this interpretability problem by using model-agnostic local explanations~\cite{ribeiro2016modelagnostic} originally designed for classifiers and adapt them to the problem of explaining learned rankings.  
We leverage the trained models and their computed measures of fit to explain the models' local behaviors using \texttt{TRIVEA} ~(Figure~\ref{figs:overview}). Transparency is achieved by leveraging visualizations that help end users generate attribute-focused, post hoc inferences~\cite{mythos} about local ranking neighborhoods.



We enable user-initiated exploration of model explanations by designing and developing \texttt{TRIVEA}, a visual analytic system that facilitates linked exploration of the goodness of fit of the models and local explanations. \texttt{TRIVEA} ensures that end users have the agency~\cite{heer2019agency,shneiderman2022human} to reason about rankings without the need to open the machine-learning black box and understand how attributes contribute to the differences between high and low-ranked entities. We conceptualized and developed the analytical and visualization components of \texttt{TRIVEA}~(Figure 1) in collaboration with researchers in machine learning, human-computer interaction, and domain sciences, such as cyber security and energy, where trust-augmented interpretation of learned rankings are a key focus area. Using \texttt{TRIVEA}, rankees and decision-makers can develop confidence in the model outcomes and build a mental model about the reasoning behind ranking differences across data subsets of interest by probing the explanations. 

As part of the conceptualization, design, and development of \texttt{TRIVEA}, we make three key contributions in this paper:
i) a principled analytical abstraction for modeling ranking labels from attributes and adopting black-box model explanation methods such as LIME for enabling the interpretability of local model behavior.
ii) design of expressive visualizations that help express model fit together with explanations comprising significant correlations among essential attributes and rankings.
iii) development of a web-based interactive system for post hoc analysis of model outcomes and explanations, the efficacy of which is demonstrated through usage scenarios and subjective feedback from a diverse group of domain experts. 

%% file: sections/relatedWork.tex
\section{Related Work}

We discuss the related work in the context of visual analytic techniques for exploring rankings and those for post hoc model explanations.

\begin{figure*}
\begin{center}
\includegraphics[width=0.9\textwidth]{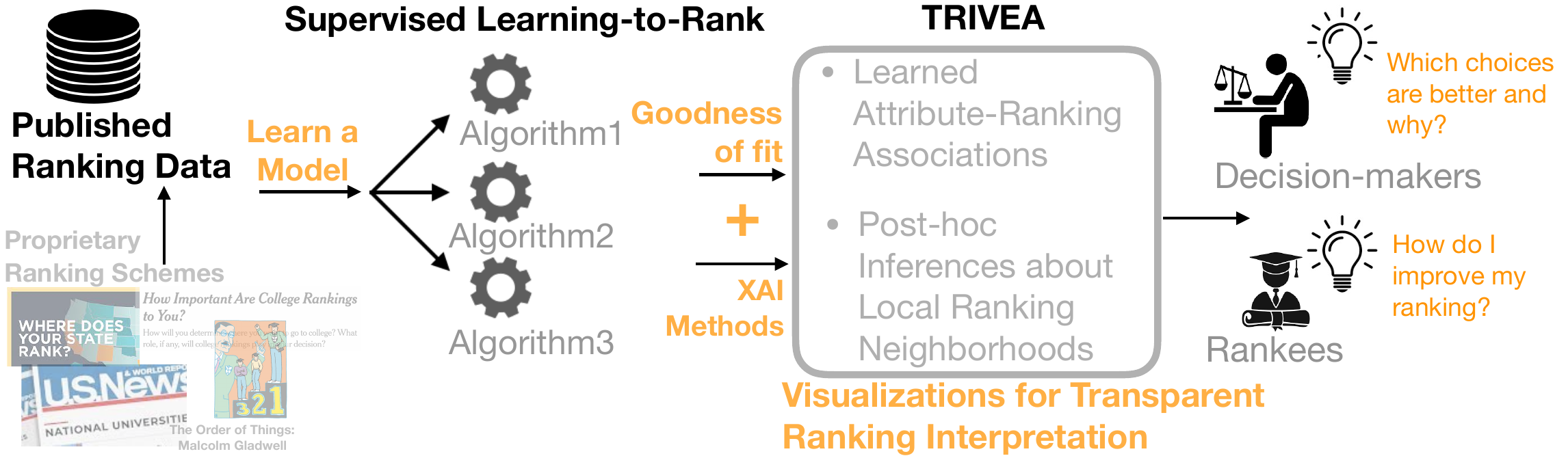}
\caption{\label{figs:overview}
{\textbf{
We address the socio-technical problem of proprietary, inaccessible ranking schemes} by using supervised learning-to-rank models that learn from published rankings and available data. 
The learned attribute-ranking associations are communicated to end users, like rankees and decisionmakers, using a visual analytic system, \texttt{TRIVEA}, that combines the goodness of model fit measures, XAI methods, and a set of interactive visualizations.
Using \texttt{TRIVEA}, rankees can generate informed, post hoc inferences about improving their rankings and decision-makers can carefully compare and contrast their choices against competing alternatives.}}
\end{center}
\end{figure*}

\subsection{Visual Analytic Techniques for Exploring Rankings}

Ranking is a convenient abstraction for human observers to quickly identify data items that can be classified into "good" or "bad" categories. 
Several visualization techniques have been proposed as a cognitive aid for 
navigating high-dimensional data spaces~\cite{seo2005rank,seo2006knowledge}, 
constructing ranking scores~\cite{perin2014table,gratzl2013lineup}, and also 
understanding changing ranking behavior across items or over time~\cite{shi2012rankexplorer}. 
Seo and Shneiderman proposed a rank-by-feature framework where ordered bars were used to guide users' attention toward high or low-ranked data items, along with a summary ranking score~\cite{seo2005rank}. 
Shi et al. used a combination of Themeriver~\cite{havre2000themeriver} and glyph-based design for showing ranking changes over time~\cite{shi2012rankexplorer}, focused on the goal of scaling visualization to thousands of items. 
A similar goal was achieved by Miranda et al., who proposed a data cube-based abstraction for efficient exploration of top-ranked data objects~\cite{miranda2017topkube}. 
For understanding score-based rankers, LineUp~\cite{gratzl2013lineup} uses a combination of stacked bar charts and interactive user assignment of weights to help users, such as university students, decide which universities could be a better choice based on their preferences.  
In all these cases, the ranking outcome is a product of human feedback or a pre-computed combination of weights with which a user can interact. In such cases, the logic of the algorithmic ranker(i.e., the scoring function) is fully accessible to the ranking users.

We focus on problems where the ranking logic or scheme is inaccessible. In that case, our approach is to 
build algorithmic rankers that model the association between attributes and rankings. Therefore, there is also a need to communicate the goodness of fit of these models so that end users can rely on them to explain the learned rankings. 

Machine learning approaches for modeling rankings have been used in Podium~\cite{wall2017podium}, which takes the user's preference of a few pairwise comparisons as training data to generate a ranking for the entire list using RankingSVM~\cite{joachims2002optimizing}. 
Podium allows users to provide a few comparisons of higher or lower-rank candidates and interactively learns the user-generated ranking by modeling a small amount of user input. In contrast, we take a supervised learning approach, allowing the algorithms to model the entire ranking using ground truth data from published rankings~(e.g., university ranking from the previous year). We use the goodness of fit measures and visualizations that communicate if learned rankings are reliable. These learned rankings ultimately serve as a means to generate post hoc inferences from visual explanations that help end users interpret attribute influence on rank positions.

\subsection{Visualization for Post Hoc Model Explanation}

We use explanation methods for black-box interpretation of machine learning models, specifically, learning-to-rank models. 
We use visualization techniques to interpret and explore the associations among data inputs and ranker outcomes. Several researchers have explored this space where black-box methods have been used for eliciting particular responses from a model~\cite{ma2019explaining,krause2017workflow,gomez2021advice,chan2020subplex,xu2022visual} from an end user's perspective or diagnosing the accuracy of classifiers~\cite{gleicher2020boxer,zhang2018manifold,arendt2020crosscheck} from a model developer's perspective. 
One of the key contributions of our work is to adopt explanation methods like LIME~\cite{ribeiro2016should}, originally developed for classifiers, to communicate explanations about learned rankings. 

However, the output from XAI methods is limited to what machine can produce and perceive the model behavior, which may not originate from a human-centered design.
We provide users the agency to create alternative groupings and observe data attribute signatures that serve as the explanation for a group of ranked items. As observed by Chan et al.~\cite{chan2020subplex}, although model interpretation at the individual level is useful, there are several visual analytic challenges for communicating group signatures. 
It is crucial to determine an aggregation scheme that is reasonable for tasks and decision-makers. Therefore, we designed a flexible and intuitive aggregation for local ranges based on the attribute's average importance and dynamic visual anchoring for aggregating explanations from multiple rankers.

For achieving these tasks, we use visual comparison methods for aiding in the navigation of ranker outputs, which has been identified as a key gap in the literature~\cite{hong2020human}.
Gleicher~\cite{gleicher2017considerations} considered the relationship between the comparison target and the action, the challenges under scalability and complexity, and the visual strategies to solve the challenges that were applied for the climate model evaluation~\cite{dasgupta2019separating}. By using a combination of visual cues and animation-based interaction in \texttt{TRIVEA}, we communicate how rankings are affected by changes in attribute importance levels.

%% file: sections/problemTask.tex
\section{Analytical Abstraction}
\label{sec:analytical}

A rank designer creates the ranking with attributes they consider important and the formula they consider reasonable.
The designer publishes the ranking and often only some attribute data and the formula.

\jun{Despite the need for rank designers to publish all data and formulas for total transparency, for a plethora of published rankings, the ranking schemes are proprietary and hence, inaccessible to the public.}
However, transparency can be increased~\jun{\cite{yuan2022rankers}} by modeling the ranking with accessible attribute data and enabling inference generation using visualizations to communicate the modeled associations.

In this section, we discuss the rationale of each step in our analytical workflow~(Figure~\ref{figs:overview}) that helps achieve such transparency.


\subsection{Problem Formulation}

\jun{
We define the following notations to formulate the problem.
The input data of an algorithmic ranker is a matrix $X$ with $n$ rows and $p$ columns.
A set of $n$ candidates or items to be ranked~(whom we term as rankees) are described with a collection of $p$ features or attributes $\{X_j\}, j = 1,2, \cdots, p$.
For a candidate $i$, its attribute values are represented as a row vector $X_i = [X_{i1}, X_{i2}, \cdots, X_{ip}]$.
An algorithmic ranker consists of a scoring formula $f(\cdot)$ and a ranking formula $r(\cdot)$.
$f(\cdot)$ receives $X$ as input, and outputs a score vector $s$.
$r(\cdot)$ receives $s$ as input, and outputs the rank vector or ranking $\tau$.
The score and ranking for a candidate $i$ are represented as $s_i$ and $\tau_i$.
The explanation about the attribute importance of candidate $i$ in ranking $\tau$ is denoted as $E(i, \tau, X)$.
We purposely do not define $E$ based only on $\tau_i$ and $X_i$ since even a single candidate's explanation is dependent on the entire ranking and attribute input.
In this work, we consider $f(\cdot)$ inaccessible, and we can only reverse-engineer or learn $f(\cdot)$ from $X$ and $\tau$.
The result of such reverse engineering is $\hat{f}(\cdot)$. 
According to different methods of reverse engineering, we may obtain multiple $\hat{f}(\cdot)_{l}$ and proxy ranking $\hat{\tau}_{l}$, $l = 1,2, \cdots, m$.
Candidate $i$'s explanation based on $\hat{f}(\cdot)_{l}$ is denoted as $\hat{E}(i, \hat{\tau}, X)_{l}$.
We want to highlight that $\hat{f}$ is not technically learning the scoring function $f$ since we only have access to $\tau$ but not the scoring output $s$.
Hence, $\hat{f}$ mimics the mixed effect of the scoring function $f$ and ranking function $r$ together.
We identify the following questions that ranking users might ask to motivate our proposed analytical abstraction:
\textbf{Q1:} Which attributes have a strong influence on the ranking, and why?
\textbf{Q2:} Does one attribute have a stronger influence on the ranking than another in local neighborhoods, and why?
In this work, we consider \textbf{Q2} as a generalization of \textbf{Q1} since we allow users to expand the ``neighborhood'' to the entire ranking range or narrow it down to a single candidate.
}

To answer the questions, we cannot simply use the ground truth rankings given by any ranking publisher. 
If we only rely on the ground truth data to understand the relationship between ranking $\tau$ and attributes $X$, 
we may use a scatter plot in which the x-axis and y-axis are $\tau$ and $X_j$.
We can observe the trend of the dots in the scatter plot to get a sense of either positive, negative, or no correlation between the ranking  $\tau$ and the attribute $X_j$.
However, such an approach cannot answer Q1 or Q2.
The alternative is to apply a trend line on the scatter plot between $\tau$ and each $X_j$. A steeper trend line indicates a stronger correlation. This approach is equivalent to applying a linear regression model between pairs of attribute $X_j$ and the ranking $\tau$ and comparing each pair's regression coefficient. 
A step further would be using a multi-variate linear regression model between attributes $X$ and ranking $\tau$. Moreover, the regression coefficients, or attribute weights, can infer which attribute has a stronger correlation to the ranking.
The inference from linear regression is easy to interpret and familiar to the public due to the long history of statistical modeling. 
\jun{
However, the algorithm may not be suitable for modeling rankings since even a ranker defined by a linear scoring function produces a ranking that is non-linear to the attribute inputs. 
Although one linear regression is not a feasible approach to generate ranker explanations, a carefully constructed collection of local linear regressions is more capable of describing non-linear behaviors.
Our choice of explanation method, LIME, is one such approach. 
It leads to opportunities to answer(Q2). But the basic linear regression can only answer (Q1).
}

\subsection{Generating Learned Rankings}


In our work, we use machine-learned rankings instead of simpler models like linear regression.
Why is linear regression not suitable for this task despite being more interpretable? 
Rankings are not continuous but integer or ordinal numbers.
It is not a common response variable handled by statistical modeling (e.g., linear regression).
Hence, although one can fit ranking with linear regression, the assumption of ranking being a continuous variable may be questionable.
Algorithms with more appropriate assumptions for ranking are being actively developed in the field of information retrieval and are commonly referred to as Learning-to-rank~(LTR) algorithms.
A common case of information retrieval~\cite{page1999pagerank} is to rank a group of webpages, so the most relevant webpages are shown at the top of the search result.
But sometimes, users click many links to find the most relevant webpage.
Researchers developed LTR algorithms to model user-perceived rankings.
In our case, the ranking data, not the scheme or the formula, is provided by the publishers (e.g., the Times University ranking~\cite{University}).
The LTR algorithms have been widely adopted outside the Information Retrieval field~\cite{liu2015application,mohler2020learning} but need to be explored more in the visualization field.
Often, a ranking publisher produces a ranking yearly, which provides multiple rankings and more rank candidate data for model training.
The multiple rankings provided by the same rank publisher across years may be considered repeated experiments, which is a desirable trait for
training the LTR algorithms.
In this work, we use the LTR algorithms to create a collection of algorithmic rankers $\hat{f}$,  produce the corresponding learned ranking $\hat{\tau}$  across years from one publisher, with the ultimate goal of modeling the influence of attributes $X$ on the ranking $\tau$.

\subsection{Explaining attribute-rank associations}

We use posthoc explanations for answering Q2 and thereby address the general need for understanding the local behaviors of the model.
Many explainable AI~(XAI) methods have been developed for quantifying the local behaviors of models. 
The two classic algorithms to explain model local behavior are LIME~\cite{ribeiro2016should} and SHAP~\cite{lundberg2017unified}.
We choose LIME because of its grounding in local linear regression. 
Each explanation from LIME can be understood as the regression coefficient from a local linear regression.
LIME is model-agnostic, which allows us to explain LTR algorithms with different flexibility and complexity to model the ranking.

\jun{The perturbation-based methods like LIME or SHAP raise concerns that the produced explanation may rely on the effectiveness of perturbation. 
In practice, there is no guarantee that more extensive perturbation leads to better explanations. And perturbation is computationally expensive.
Another group of explanation methods based on partial dependency plot~(PDP) and individual conditional expectation~(ICE)~\cite{yeh2022bringing} does not rely on the perturbation of individual data points and thus is relatively less computationally expensive. Keeping the pre-processing steps the same, we compare the degree of agreement between alternative explanation methods and allow end users to visualize such comparisons.}



\par \noindent \textbf{LIME explanations:} LIME summarizes the correlation between the ranker input and output via perturbation on the input based on the distribution of the background data. We set the background data to be all the data points across the years, comprising published rankings. 
The raw output of a ranker is numerical ranking scores. 
One can convert the scores to a ranking by sorting the scores, typically in descending order. 
We chose to use the ranking scores instead of the ranking for LIME to generate inferences since using ranking may result in a sampling imbalance. For example, when LIME applies perturbation on the ranking at rank position one, the scores can increase or decrease. However, the rank can either stay at rank one or decrease, so LIME cannot effectively derive the correlation between attribute and ranking. Also, ranking scores are a direct indicator of the ranker's behavior. 

LIME ignores attribute dependence, which, in our case, can lead to negative contributions that are counter-intuitive. For example, an attribute that is supposed to have a positive contribution as a rank stimulator, but is not as effective as another dependent attribute, results in a negative regression coefficient. Since our goal is to use explanations as decision-making aids for lay users, we choose to normalize the contribution between $0$ and $1$ per ranker.
For rankers, such normalization retains the relative difference between the attributes, which does not interfere with the comparison of the attribute importance.
The alternative is to force LIME to produce non-negative contributions, but that would affect the explanation quality and will not solve the attribute dependency issue, which is out of the scope of this work.
LIME generates explanations for each rank candidate, 
which allows us to group and compare them within and across rank ranges.
A key contribution of this work is to adapt the LIME output and use interactive visualization to support users in making post hoc inferences about local rank neighborhoods.

\jun{
\par \noindent \textbf{Gauging agreement between LIME and ICE feature impact:}
In this work, we adopted the ICE feature impact explanation and compared it with the LIME explanation output. 
We analyzed the similarity between the ICE and LIME explanation using Pearson correlation. 
We observed that, for some rankers, when the ranker produced rank was closer to ground truth, the explanations methods had a greater degree of agreement between them. 
However, this was not consistent for all rankers or all rank ranges, and hence we deemed it judicious to leave it to the end user's judgment for the choice of an explanation method. The original ICE feature impact paper~\cite{yeh2022bringing} averages the ICE feature impact of all instances to obtain a single overall impact score for a certain attribute. 
We rewrote the equations~(and the code) to seamlessly fulfill instance-wise and group-wise feature impact calculations, similar to the pre-processing steps for generating LIME explanations.
In this way, the ICE explanation data structure is aligned with LIME, leading to easier computational comparison and a unified user interface back-end.
}

\begin{table}[tb]
  \caption{Evaluation Metrics for Trained Algorithmic Rankers}
  \label{tab:metrics}
  \scriptsize%
	\centering%
  \begin{tabu}{cccc}
\toprule
ranking data   & algorithm   & NDCG@10 & P@10 \\
\midrule
University  
  & Cord.Ascent & 0.20    & 0.07 \\
  & \textbf{LambdaMART}  & 0.64    & 0.98 \\
  & ListNet     & 0.19    & 0.08 \\
  & MART        & 0.56    & 0.87 \\
  & RankBoost   & 0.48    & 0.75 \\
  & \jun{RankingSVM}   & 0.65 & 0.97 \\
  
\midrule
Fiscal
 & Cord.Ascent & 0.35    & 0.32 \\
 & LambdaMART  & 0.38    & 0.42 \\
 & ListNet     & 0.39    & 0.55 \\
 & \textbf{MART  }      & 0.63    & 0.95 \\
 & RankBoost   & 0.47    & 0.67 \\
 & \jun{RankingSVM }  & 0.52 & 0.87\\
  \bottomrule
  \end{tabu}%
\\
Metric scores are between $0$~(worst) and $1$~(best).
\end{table}
\jun{We train LTR models a collection of $\hat{f}$ from the RankLib package~\cite{RankLib}. We compare their performance with the ranking SVM model that is implemented according to the Podium paper~\cite{wall2017podium} (in Python), and explain them using the LIME~\cite{LimeRepo} Python project and ICE Python project~\cite{IceRepo}. We import models' output $\hat{\tau}$ and corresponding inferences $\hat{E}(i, \hat{\tau}, {X})$ into \texttt{TRIVEA}.}

\begin{figure*}
\begin{center}
\includegraphics[width=0.98\textwidth]{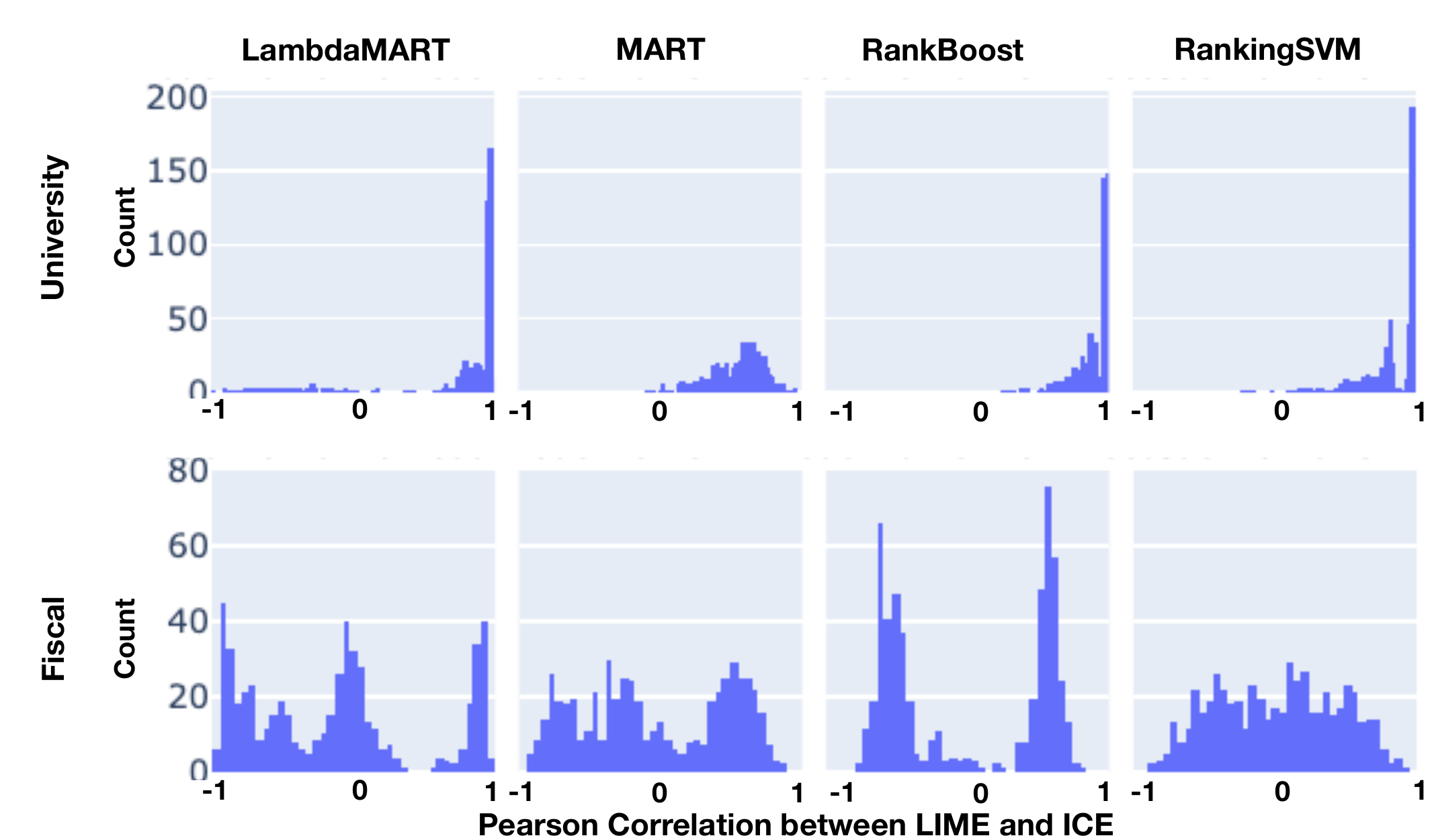}
\jun{
\caption{\label{fig:explainCompare} 
\textbf{Evaluation of explanation agreement for the most accurate models from Table 1.} The x-axis is the item-wise Pearson correlation between LIME and ICE explanations. A higher value indicates more agreement between explanations for a ranked data item. The y-axis is the count of the data items.
}
}
\end{center}
\end{figure*}

\subsection{Comparison of Model Performance and Explanations}

\jun{We use the university rankings data~\cite{University} and the state fiscal rankings data~\cite{Fiscal} for the two scenarios. The LTR models trained in the scenarios have been evaluated with the traditional metrics, like Normalized Discounted Cumulative gain~($NDCG@10$) and Mean Average Precision~($Precision@10$) as shown in Table~\ref{tab:metrics}. 
For both metrics, the range is 0 to 1, and a model with a score closer to 1 is deemed to be more accurate. Table~\ref{tab:metrics} shows that a model high on one metric may be low on the other. 
Also, based on the average across both metrics, the best models for university data are LambdaMART and Ranking SVM; and MART for Fiscal data.}

\jun{
We analyzed the instance-wise agreements between LIME and ICE explanations in Figure~\ref{fig:explainCompare}.
It shows that the agreement between LIME and ICE explainers differ across rankers.
Additionally, the ranker shows different distributions between the University and Fiscal data.
Overall, a high degree of agreement is indicated for the University data set, as most of the plots show left-tail distribution, indicating that for most ranked items, the Pearson correlation between LIME and ICE explanations is high.
On the other hand, for the Fiscal data, we show that many rankers exhibit a uniform or multi-modal distribution, indicating lesser agreement. 
We conclude that the agreement among explanations is subject to conditions~(e.g.,  ranker, data properties, etc.) and should be left to the judgment and interpretation by end-users. 
Hence, we inject transparency into the ranker interpretation process by designing an interactive user interface as part of \texttt{TRIVEA} that seamlessly provides information about learned rankings and corresponding explanations from multiple rankers and explanation methods.
}



\subsection{Measuring goodness of fit}

Measures of goodness of fit can express model uncertainty~\cite{pang1997approaches,maack2022uncertainty} and can be described as the deviation between output $\hat{\tau}$ to the ground truth ranking $\tau$.
A smaller deviation indicates better goodness of fit.
More flexible algorithms tend to have better goodness of fit when modeling complex relations between attributes $X$ and ground truth ranking $\tau$. 
The standard evaluation metrics for LTR models such as NDCG~\cite{valizadegan2009learning} and MAP~\cite{robertson2008new} are designed for better webpage ranking or Information Retrieval ranking in general.
For instance, NDCG, or Normalized Discounted Accumulated Gain, measures the goodness of fit of top-ranked webpages with an exponentially larger weight than the lower-ranked webpages' fit.
However, users may be more focused on the goodness of fit in a certain range other than the top.

The standard evaluation metrics~(Table~\ref{tab:metrics}) or other summary metrics do not capture local model behavior. For example, a model A that has a greater average precision score than model B, might have errors in local neighborhoods that a user might care about. Hence, we need local and granular measures of goodness of fit. 
One of the contributions of this work is to adapt the deviation between model output ranking $\hat{\tau}$ to the ground truth ranking $\tau$ and visualize the goodness of fit interactively. \jun{By interactively visualizing both goodness of fit and the LIME and ICE explanations, end users can transparently gauge model uncertainty and whether to trust an explanation given the degree of deviation between the ground truth and the learned ranking.}

%% file: sections/design.tex
\section{TRIVEA: Tasks and Interface Design}

We designed a web-based visual analytic system as part of \texttt{TRIVEA} for facilitating learned ranking-driven inferences. By enabling post hoc interpretation and reasoning about the behavior of multiple models.  Rankees, like university administrators, can try and understand competitors' characteristics and compare them with their own for improvement. On the other hand, decision-makers, like students or stock market investors, can draw inferences from published rankings and the associated attributes to drive their future investment~(i.e., educational or financial, respectively) decisions.
In this section, we outline the tasks and design rationale of our interface that guides the organization of the interface components. We confirmed the ecological validity of the tasks and the relevant design rationale through discussions and pilot studies with four data science practitioners in the industry. By demonstrating intermediate prototypes in the pilot studies and collecting their design feedback, we refined the tasks and visualization design realized in \texttt{TRIVEA}.

\subsection{Visual Analytic Tasks}

After deriving the analytical abstraction(Section~\ref{sec:analytical}) we focused on visualization interventions for communicating the goodness of fit of alternative algorithmic rankers and their explanation, as well as for allowing end users rich interactivity for exploring local ranking neighborhoods. 
We derive the following visual estimation and interpretation tasks accordingly: i) {Estimate local goodness of fit of rankers }(\textbf{T1}): As part of this task, our focus is on detecting the discrepancy between the learned ranking~($\hat{\tau}$) and ground truth~(${\tau}$) for each data item. Global metrics such as mean average precision~\cite{robertson2008new} cannot capture discrepancy item-wise. Therefore, we use these metrics as a guide for automatically suggesting models or rankers with high accuracy (e.g., mean average precision is 1) but use visualizations to communicate itemized discrepancy. 
ii) Understand attribute importance in local rank neighborhoods~(\textbf{T2}): As part of this interpretation task, our focus is on efficiently communicating the relative importance of attributes on rankings in local neighborhoods using the LIME explanations~($\hat{E}(i, \hat{\tau}, {X})$ ), and 
iii) Detect correlation between attribute values and importance~(\textbf{T3}): This task entails a more detailed inspection upon observation of relative attribute importance. Taken together, T2 and T3 help gather evidence for generating inferences about what contributes to rank with respect to any specified groupings (e.g., subset by attributes values, subset by attribute contribution values). 
\jun{T2 results in observations that help to answer the first part of Q2~(i.e., Does one attribute have a stronger influence on the ranking than another in local neighborhoods?), and T3 results in inferences that help to answer the second part~(i.e., Why the influence differ?). T1 helps to estimate the credibility of the observation and the inferences from T2 and T3. All tasks can be applied to an arbitrary size of a local neighborhood, hence Q1 can be answered as well.}

\subsection{Interface Overview and Design Rationale}

\begin{figure*}[t]
\begin{center}
\includegraphics[width=\textwidth]{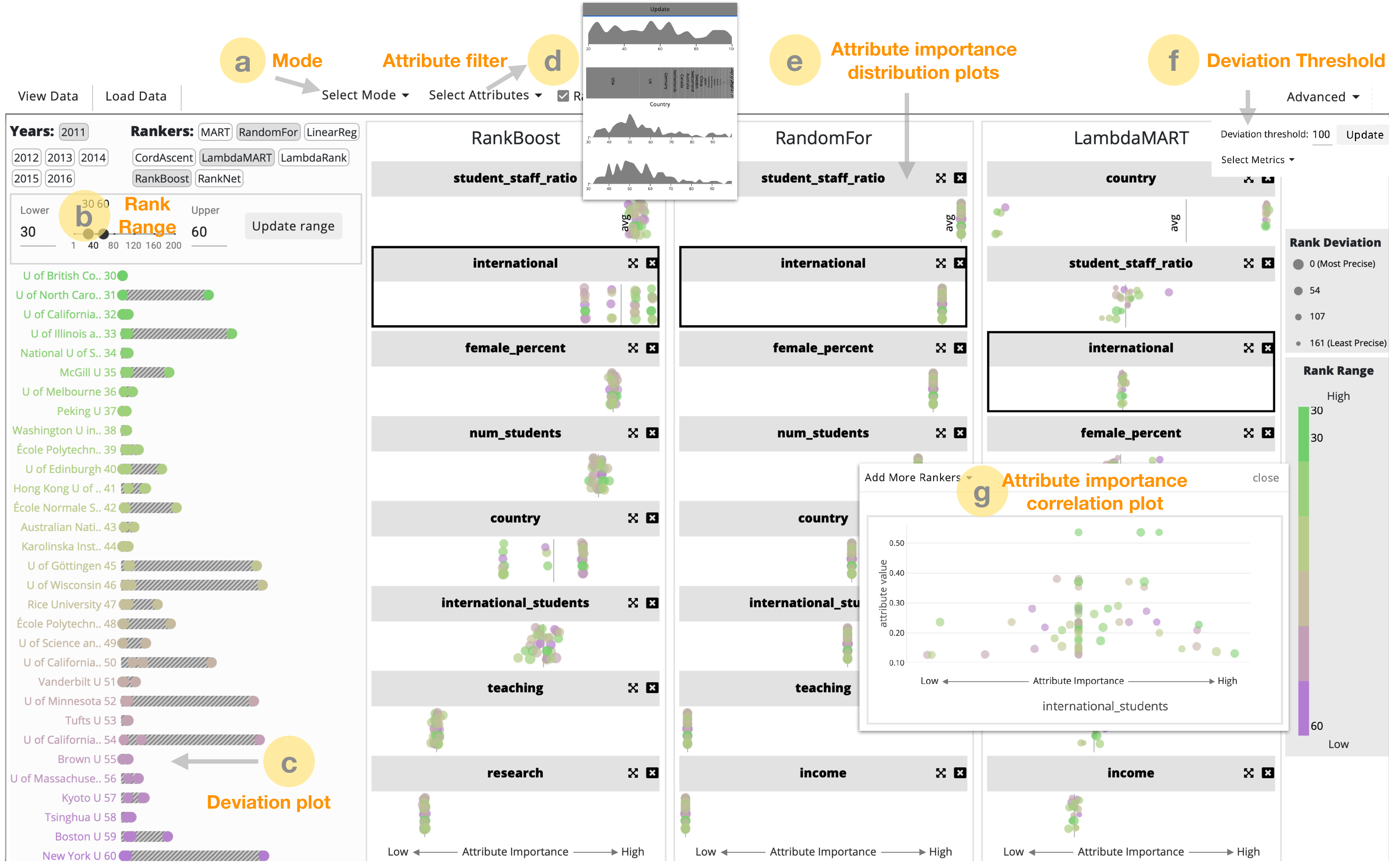}
\caption{\label{fig:interface}
\textbf{Visualization Interface for \texttt{TRIVEA}}: The control panel on the left comprises options to select:
(a) different modes of comparison, such as Ranker mode, Range mode, and Time mode,
(b) a group of items by rank range,
(d) a group of items by  attribute values, and 
(f) the permissible deviation threshold. 
The visualizations shown are: the \textbf{deviation plot}~(c), which encodes the goodness of fit for the learned rankings; 
\textbf{attribute importance distribution plots}~(e), which shows the attribute importance distribution~(x-axis) among the items in the selected range, \textbf{attribute importance correlation plots}~(g), which shows the attribute importance~(x-axis) versus the ground truth value~(y-axis).
In (e), we add \textbf{jittering along the y-axis} to minimize overlapping. The Y-axis does not carry any meaning. The attributes are \textbf{sorted} from top to bottom according to the attribute average importance score for the given range and ranker.
}
\end{center}
\end{figure*}

\texttt{TRIVEA}~(Figure~\ref{fig:interface}) consists of the following components: a control panel for user selection of instances and attributes, based on rank-range and attribute ranges, respectively; 
a set of filters for sub-setting across models, and data items, attributes, or the year of interest;
and visualizations such as the deviation plot~(Figure \ref{fig:interface}c), attribute importance distribution and correlation plots~(Figure \ref{fig:interface}e, g). We discuss the interface components below in the context of the relevant design rationale for realizing the tasks T1, T2, and T3.

\par\noindent \textbf{DR1: Enable multi-way visual comparison:} For satisfying T1, T2, and T3, we want to create appropriate interaction affordances for quick user navigation of the data space based on items and attributes of interest and comparing rankings by understanding the algorithmic goodness of fit and reasons behind model outputs.
Comparison tasks can be expensive in terms of the amount of human attention required to separate signals from the noise caused by clutter or irrelevant information. We link model outcomes and explanations using colors that encode rank positions. We use a diverging color scheme for the chosen rank range of data items displayed, which helps add contrast between high and low-ranked items within the local range.  
To allow flexible comparison, \texttt{TRIVEA}  has multiple modes of comparison~(Figure \ref{fig:interface}a):
a) Ranker mode: one can compare across multiple rankers, b) Range mode: one can compare between different rank ranges for a given ranker, and c) Time mode: one can compare between different years for a given ranker and rank range.
We use linked views, where ranking positions in local neighborhoods need to be associated with attributes that are considered important for the model outputs. One can also visually link across multiple models, as shown by the black-highlighted attributes, to observe if there is reasonable consensus about the model output and the attribute-based explanations~(T2).

\par\noindent  \textbf{DR2: Enable dynamic comparison anchoring:} Since we communicate the outcomes from algorithmic rankers ensemble, it is essential to anchor comparisons based on an end user's perspective. We could either use the model outcomes as comparison anchors or the ground truth ranks. Based on pilot studies and feedback from our collaborators, we made a deliberate design choice to anchor comparison and user navigation based on ground truth ranks. Since the model's goodness of fit is conveniently communicated across all the visualizations~(T1), we preserve the mental model of an end user who might choose data entities based on their prior knowledge~(e.g., university administrators or students who are interested in schools belonging to some known rank range) and also communicate the reliability of the model outcomes in that local rank range.
We allow users to highlight the attributes and the rankees in the interface as \textit{visual anchors}. Users can observe the rankees and attributes of interest while changing other functions. Users can adjust the rank range, tweak the deviation threshold~(Figure \ref{fig:interface}f), change the model selection, compare the current rank range to a different rank range,  compare the current ranking year to a different ranking year, etc. We use animations to guide the users' attention toward relevant changes in explanations.


\par\noindent \textbf{DR3: Provide user control for defining local groupings:} We provide users with control over which items they want to focus on, or which models they think best reflect their mental model about ranked items, while at the same time, we provide guidance to users to support their task of looking at rankings from a model's perspective. 
The data filters in \texttt{TRIVEA} can help users to stay close to their mental models about the ranked items. Users may evaluate the models based on the outputs and the attribute importance associated with the local groupings in the subset created through the data filters.
The data filters consist of: \textbf{Range selection}: In the default selection, where users may like to compare across different models, they can use the range selection filter to select a specific rank range of interest. For example, users can select rankees in the rank range of $30$ to $60$~(Figure \ref{fig:interface}b).
\textbf{Attribute selection}: Users can use the attribute selection filter~(Figure \ref{fig:interface}d) to select the items by their attribute values. For example, an analyst can select universities with a female student ratio above forty percent.



\section{Visualization Design and Interpretation}
\label{sec:visualizationsdesign}

In this section, we describe how our design choices for the interactive visualizations impact the interpretation of learned rankings in local data neighborhoods.
We use the Times Higher Education ranking~\cite{University} as a running example to explain the system component.
The data comprises $818$ unique universities from the year $2011$ to $2016$ with $10$ attributes. There are $12$ columns consisting of $1$ ranking, $1$ total score and $10$ attributes, including \textit{teaching}, \textit{teaching}, \textit{female percentage}, \textit{international student percentage}, \textit{research},etc.  
Since the ranking formula is unknown, we can use the historical university ranking data to build algorithmic rankers that approximate the original ranking formula and generate inferences between attributes and ranking.

\subsection{Understand goodness of model fit~(T1)}

\begin{figure}[t]
\begin{center}
\includegraphics[width=\columnwidth]{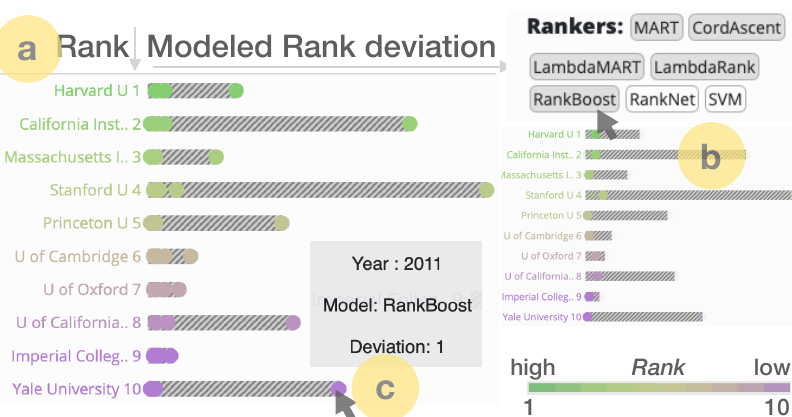}
\caption{\label{figs:deviation_plot} 
\textbf{Deviation Plot} 
a) encodes ground truth ranking and the deviation of the learned ranking from the ground truth,
b) interaction to focus on one ranker by hovering over the ranker button or 
c) focus on one ranker \& item by hovering over the dot. 
}
\end{center}
\end{figure}

We designed the {deviation plot} for visualizing item-wise goodness of fit of learned rankings. 
For addressing \textbf{T1}, we use the position channel as the primary visual cue for indicating \textit{item-wise} \textit{goodness of fit}, measured by the absolute distance between the \textit{original rank position} and \textit{modeled rank positions}. We use a striped texture as a metaphor for ``poor fit": the larger a stripe, the greater the error in the learned ranking.

As shown in Figure~ \ref{figs:deviation_plot}a, dots close to the y-axis on the left indicate a more accurate ranker. If dots from multiple models converge close to the y-axis, as for rank position 9, we can infer that most models are accurate. We can observe that for most other rank positions, there are models that are inaccurate, as indicated by dots farther away from the y-axis. To identify the name of the model, one can select the dot or highlight a ranker as shown in Figure~ \ref{figs:deviation_plot}b. Users can also hover over a dot to activate a tool-tip window describing the specific model Figure~\ref{figs:deviation_plot}c.

Note that we do not differentiate between the directions of rank position deviation. As observed in Figure~\ref{figs:deviation_plot}, the deviation plot can communicate inter-model agreement/disagreement by letting users compare multiple models' \textit{goodness of fit} with respect to the same \textit{ground truth ranking position}.

\begin{figure*}
\begin{center}
\includegraphics[width=0.97\textwidth]{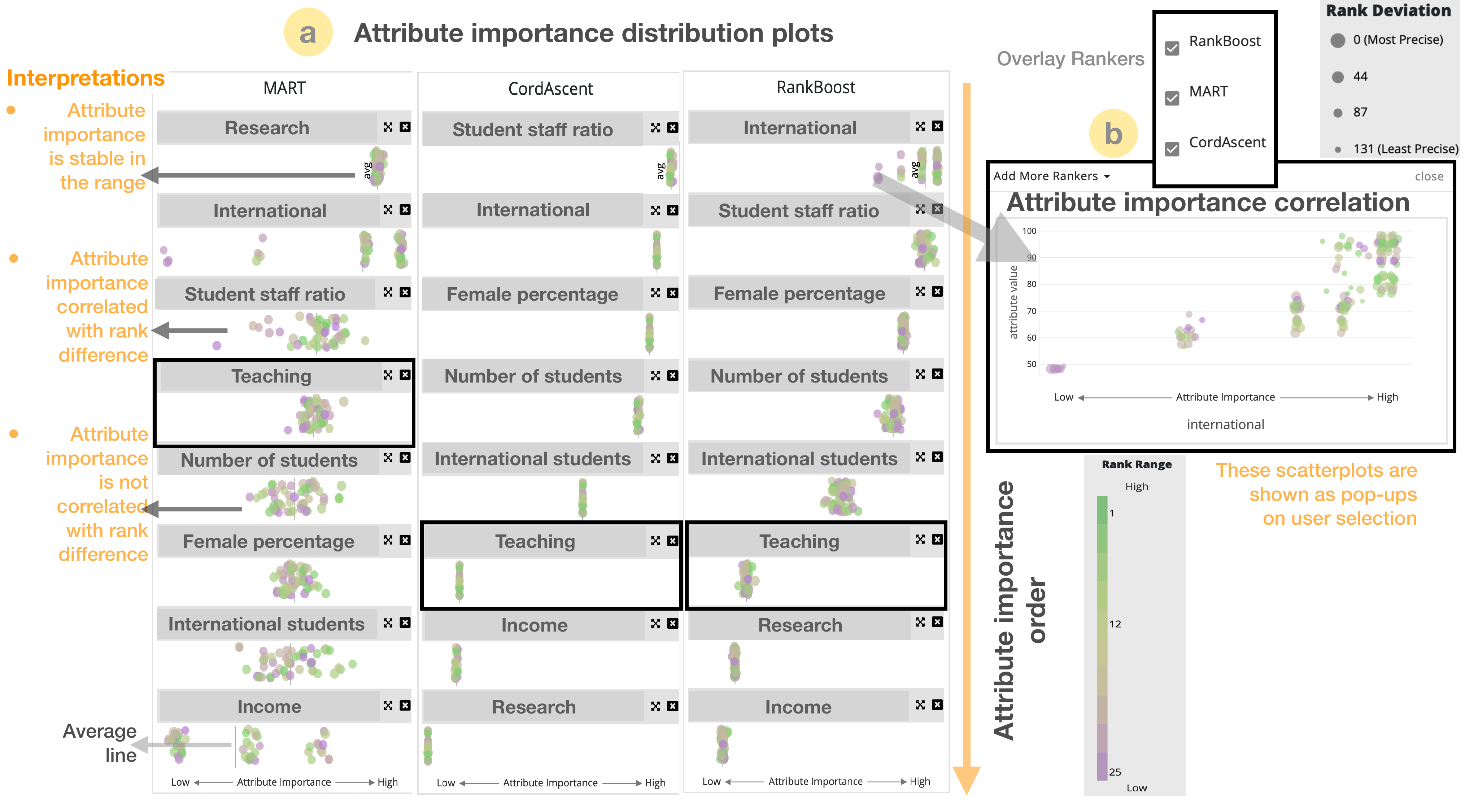}
\caption{\label{fig:explanation}\textbf{Attribute importance plot} encodes outcomes from ranker explanations and helps reveal if attributes are important to rankees in the local region, rankees compared among themselves, and the rankees' attributes value. The different components, as shown in this figure, are 
(a) Attribute importance distribution plots that show the relative attributes' importance among the rankees for each ranker,
(b) Attribute importance orders that show the attributes' relative importance for the local region for each ranker, 
(c) Attribute importance correlation plot that shows the correlation between attribute~(e.g., \textit{International}) value~(y-axis) and importance~(x-axis) for one ranker, or add more overlaid rankers for gauging their consensus.
}
\end{center}
\end{figure*}

 
\par \noindent We considered several alternatives to deviation plots, 
and evaluated intermediate prototypes through subjective feedback from our collaborators. Here, we discuss the rationale behind the selection of the deviation plot as the final design choice.
It would be hard to see the actual deviation between the ground truth and the model outputs for a single model or a group of models with scatter plots. 
With bar charts, it would be difficult to explicitly encode the relative ranking difference~(between ground truth and model output) along with the absolute value of the rank positions. 
With a heat map, while color can help spot differences quickly, it would be difficult to encode changes across years using color, leading to change blindness. However, we preserve the use of color channels by using a diverging color scheme for distinguishing the rank range of the items. A neutral color reflects the center rank for the selected rank range.
In our pilot studies, we used slope plots as an alternative. Slope plots effectively demonstrate the rank increasing and decreasing between the ground truth ranking and model output. However, the slope is sensitive to smaller rank changes and gradually less sensitive to larger ones. Also, slope plots can get cluttered when there is a high degree of discrepancy between model outcomes and ground truth ranks.
The direction of increasing or decreasing the rank is less meaningful. Deviation plots can directly express the rank deviation no matter the direction and identify the less deviated output from one model or model ensemble by letting users choose the deviation threshold as a tolerance for the goodness of fit.


\subsection{Interpreting Explanations~(T2, T3)}

Understanding the importance of attributes to the ranking is crucial for building trust and gaining insights into the instance’s attributes. In this section, we will discuss how to visualize attributes' effects on ranking. 

\par \noindent \textbf{Interpreting attribute importance order:}
LIME explanation is a score-based explanation. For example, an attribute with a high LIME score is more important than a low score. However, the scores are not the most effective way to communicate which attribute is more important, for a single rankee or a group of nearby rankees. For a group of rankees in a certain rank range, users may want to know the overall most important attribute in such a range. Hence, we take the group average of each attribute's contribution scores to sort the attributes as shown in (Figure~\ref{fig:explanation}). 

The number of attributes can vary across data sets; thus, it is essential to understand which attributes are important for generating rankings by a ranker. Hence, we have sorted the attributes based on attribute importance order.
Attribute importance order is the highest abstraction of the contribution scores, which allows users to understand the group-wise attribute importance in a nutshell.
For example, \textit{research} is a more influential attribute compared to the \textit{student staff ratio}, hence \textit{research} is shown before the \textit{student staff ratio }~(Figure \ref{fig:explanation}a, left). This will help an analyst skim through the most important attributes.
Although we, by default, allow users to see the eight most influential attributes, due to the limitation of the browser window, we provide a ``remove" bottom to eliminate an attribute from the queue. And the next most important attribute in the queue appears in the interface.  In this way, we allow users to access the entire attributes for exploration and benefit from the XAI-driven attribute importance suggestions.


\par \noindent \textbf{Interpreting attribute importance distribution:}
We designed the attribute importance distribution plot for attributes' contributions scores from one ranker (Figure~\ref{fig:explanation}a).
It consists of attribute dot plots and provides visual cues of proximity to identify distributions in the attribute space. 
Each dot plot contains an average line for each attribute aiding in the comparison across attributes. The attribute importance order is based on this average score.
We also encode the rank deviation defined in the deviation plot as the dot size. 
The larger the deviation, the smaller the dot size. So more accurate data points are more visible in the attribute importance distribution plots. 
Hence, not only the deviation plot is used to communicate the goodness of fit of the algorithmic ranker and algorithmic rankers ensemble, but attribute importance distribution plots are designed to facilitate linked comparison of the goodness of fit and explanation across multiple rankers. 
Users can filter out less accurate dots by the deviation thresholds.
By controlling the deviation thresholds and visualizing the deviation as the dot size, users are guided to pay more attention to the attribute with larger dots, indicating more reliable explanations. 
In practice, users can first understand the relative attribute importance within a local range of interest. Then, users can investigate the attributes of interest as ordered. 
Sorting the importance distribution plots by importance order is particularly useful when the number of attributes is large.

The contribution scores can have varying ranges for different attributes, making it difficult to compare the contributions across attributes and rankers. 
Hence, we have standardized the attribute contributions between $0$ to $1$ per ranker in the given rank range. 
The average reference line on the x-axis reduces the information load for individual comparisons and gives users an intuitive understanding of the relative difference in attribute contributions. It also maintains useful decision-making guidance based on the relative contribution of each attribute, such as the relative reliability or stability of the attribute importance. 

For example,  we can observe that 
in attribute importance distribution plot~(Figure ~\ref{fig:explanation}a,~left), for the attribute \textit{research}, data points are all distributed near the average, but for \textit{international}, they are distributed across the range. This means \textit{research}'s importance is more stable than \textit{international} in the given rank range, but \textit{international}'s importance varies across rankees, and they also appear to be clustered at different rank positions.
An analyst can also observe that the dot size encodes the rank deviation~(i.e., the larger the deviation, the smaller the dot size). 
Not only the rank deviation links the attribute importance distribution plot and the deviation plot, but also the color that encodes the relative rank positions in a given rank range.
But based on the color of the dots, we cannot tell if the \textit{international}'s importance correlates to relative rank positions.
But for \textit{Student staff ratio}, the green dots are mostly on the right side of the average and purple on the left. That means the \textit{Student staff ratio} is more important for higher-ranked rankees in the given rank range.
On the other hand, for \textit{Number of students}, we observe that the dots are randomly scattered with no correlation pattern between the color of the dots and their respective importance scores.


\par \noindent \textbf{Interpreting correlation between attribute importance and attribute value:}
Users may want to compare the attribute value and contribution in one ranker or across several rankers to understand inferences about attribute value and attribute importance. 
The attribute importance correlation plot~(Figure~\ref{fig:explanation}b) shows the attribute's contribution to the ranking on the x-axis and the attribute value on the y-axis. The attribute importance correlation plot inherits the dot size encoding from the attribute importance distribution plot so that users can generate inferences based on better goodness of fit.
In the attribute importance correlation plot for \textit{international}, for each point, the y-axis is the international score for a school, and the x-axis is the contribution of a such international score for that school's ranking. Users can choose to see a specific ranker to understand how such ranker regards this attribute's importance or let multiple rankers overlay their points on one plot to understand the consensus among the rankers. For example, if a school's international score is $30$ on the y-axis, and the user wants to see the five rankers' consensus on an international score of $30$ in the given rank range. In that case, five points represent five different rankers, having the same y-axis value of $30$, but different x-axis values according to their contribution scores.  

As shown in Figure~\ref{fig:explanation}b, this popup view shows the attribute importance correlation plot with an explicit y-axis so that the users can refer to the actual attribute values.
We expand each attribute's importance to the entire x-axis range. This will help an analyst understand the attributes' values that have relatively strong or weak importance for an individual ranker or among rankers.
For example, if a university administrator observes that lower international scores show weaker importance for a certain ranker, they can improve the international student percentage during their admission process. But what if the administrator wants to understand the multi-rankers consensus around this attribute, i.e., if this same attribute's importance is similar across other models? This task can be achieved using the ``Add more Ranker" option. Selecting different rankers from this option will plot the attribute contribution scores of each of those rankers. It helps the user compare and understand the attribute importance for several rankers altogether. Through the comparison, users can understand ranker consensus on the particular attribute. 
The inferences derived in the attribute importance correlation plots are actionable. An example of such an inference would be: increasing international student percentage or international collaboration can promote the rank of certain schools in the rank range. Using \texttt{TRIVEA},  one may derive multiple such inferences and use their domain knowledge to determine which inferences are actionable.

%% file: sections/results.tex
\section{Usage Scenarios}

In this section, we discuss two usage scenarios using \texttt{TRIVEA}. We use the university rankings data~\cite{University} and the state fiscal rankings data~\cite{Fiscal} for the two scenarios.
The models and explanations are first generated and imported into the interface.




\subsection{Understanding States' Fiscal Ranking Change}
\label{sec:state}

We present a usage scenario demonstrating how \texttt{TRIVEA} 
can be used by state administrators for interpreting and acting upon state fiscal rankings and their explanations.
\label{sec:usage_scenario_fiscal}
The state fiscal data set~\cite{Fiscal} comprises rankings of $50$ U.S. states with $33$ attributes about the financial performance of each state from year $2006$ to $2016$. The attributes include state financial metrics such as\textit{ primary government debt}, \textit{total net asset},\textit{ cash ratio}, \textit{tax income ratio}, and more.
The rankings depict the financial status of each U.S. state and are generated using a formula that the formula maker predefines to produce the ranking. However, the entire methodology of making such a formula is complicated and confusing. How can a state administrator judge if the formula is reasonable without knowing the formula? Especially since every formula is a simplification of the real world. Hence, the administrator can run algorithmic rankers that abstract the important attributes for ranking and see if that is reasonable. With the XAI techniques, it will also provide formula makers another perspective regarding the formulae and how well it works with the data.
We assume a scenario in which a government officer from New Jersey~(NJ) working in the finance department wants to understand why an NJ's fiscal ranking increases or decreases over the years. 

\begin{figure*}[t]
\begin{center}
\includegraphics[width=0.98\textwidth]{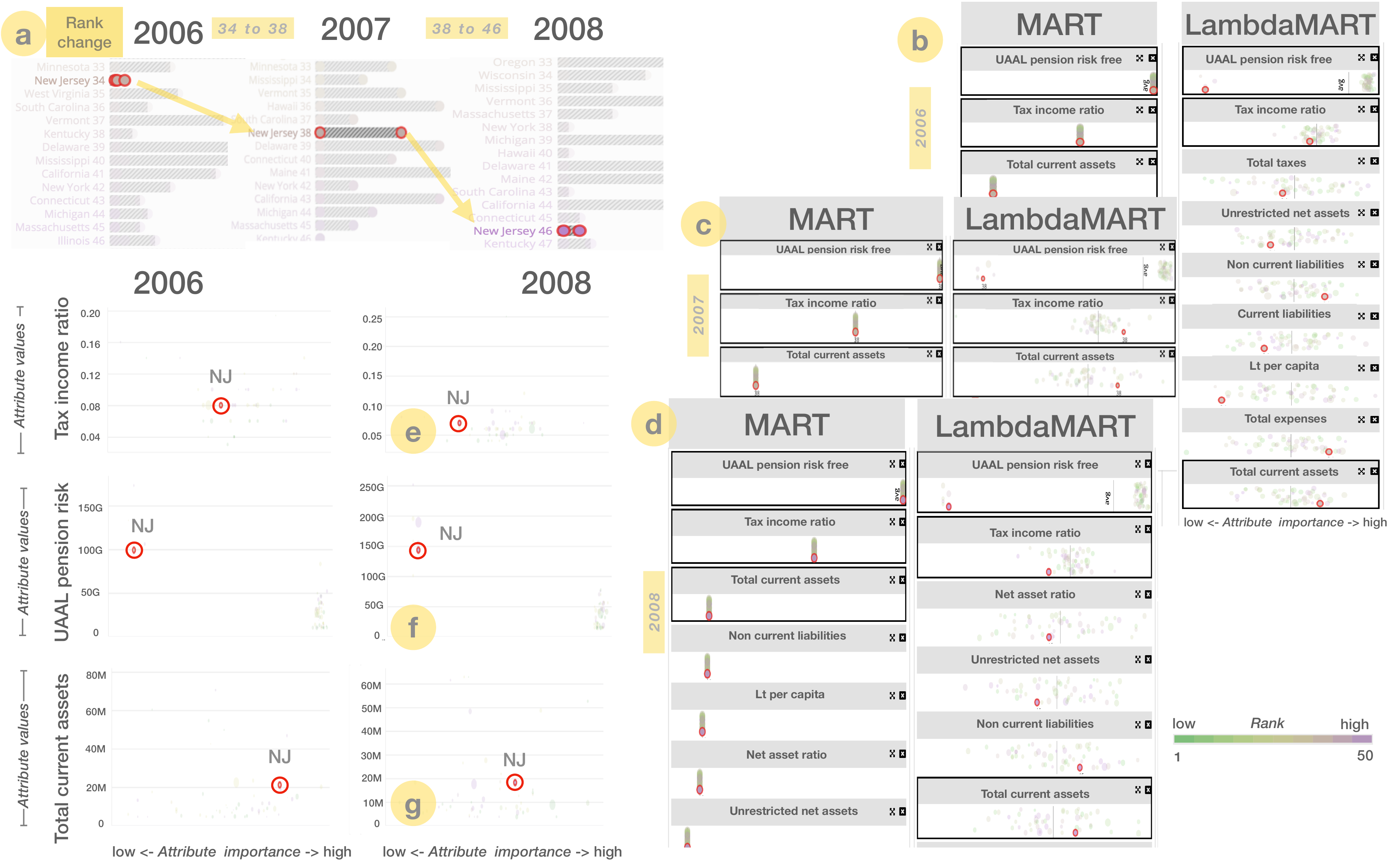}
\caption{\label{fig:usageScenario} \textbf{Usage Scenario: States' fiscal rankings~(Section~\ref{sec:state}).}
a) Deviation plot for years from 2006 to 08, with highlight on the state of NJ. The highlight is kept in the entire interface;
b), c), d) show attribute importance distribution plots for rankers in the years 2006, 2007, and 2008. Each column is sorted by the importance score average;
e), f), g) show attribute importance correlation plots between the years 2006 and 08, with highlights on the state of NJ. Here, the x-axis is the attribute importance, and the y-axis is the attribute value.
}
\end{center}
\end{figure*}

\jun{She set the year as 2006 and wanted to observe the ranker consensus for a small deviation threshold on the deviation plot. MART and LambdaMART~(Figure~\ref{fig:usageScenario}a,b) were the two best rankers~(\textbf{T1}).} 
She observed that LambdaMART's explanations~(Figure~\ref{fig:usageScenario}b,c,d), as indicated by the spread of the importance scores for the different attributes, were in contrast to MART's and could provide alternative interpretations about attribute influence on the rankings~(\textbf{T1}).

She understood that out of the 33 attributes in the data, the rankers showed consensus patterns on the top eight attributes~(\textbf{T2}) that are shown on the \texttt{TRIVEA}  interface.
She now focused on understanding the relative importance of the attributes from the rankers' explanation.
She removed the \textit{quick ratio} since it did not seem interesting to investigate right now. She highlighted the \textit{UAAL pension risk free}, \textit{tax income ratio}, \textit{total current assets}, the top three according to the MART ranker~(shown as the rectangle boxes in Figure\ref{fig:usageScenario}b), which she wanted to investigate. The three attributes also appeared on LambdaMART explanation~(shown as the rectangle boxes in Figure\ref{fig:usageScenario}b), but the relative importance of those three attributes varied across years in the top eight attributes~(\textbf{T2}). 
Therefore, she inferred that in the year 2006, the LambdaMART ranker agreed with the relative order among the top three attributes with MART ranker, but the third attribute was the eighth~(shown as the third rectangle boxes in Figure~\ref{fig:usageScenario}b) ranked for LamdaMART~(\textbf{T2}). 
At this point, she highlighted NJ~(shown as the red dot) and saw that, in LambdaMART ranker explanation, \textit{UAAL pension risk free} was not important to NJ's ranking. This meant it was not helping NJ compete with other states. The \textit{tax income ratio} had average importance compared to the average line, and \textit{total current asset} was very important. She suspected that NJ benefited more on the \textit{total current assets} than the other two attributes. The values of the three attributes compared to other states were high, middle, and high, observed from the importance correlation plot~(\textbf{T2})~(Figure~\ref{fig:usageScenario}e,f,g).

She wanted to observe the importance of the three attributes across the years. From 2006 to 2007, all three attributes' importance decreased for NJ~(Figure~\ref{fig:usageScenario}b,c). NJ's rank dropped five positions as observed in the deviation plots~(Figure~\ref{fig:usageScenario}a). 
From 2007 to 2008, the rank dropped another seven positions~(Figure~\ref{fig:usageScenario}c,d), the \textit{tax income ratio} and \textit{total current assets}' attributes' importance decreased for NJ~(Figure~\ref{fig:usageScenario}a). 
She was curious about what happened to  NJ between 2006 and 08 that led to the drop from rank 34 to 46. For NJ's attribute values, the  \textit{UAAL pension risk free} increased from 100G to 150G~(Figure~\ref{fig:usageScenario}f).  \textit{tax income ratio} increase from 0.06 to 0.07~(Figure~\ref{fig:usageScenario}e), and the  \textit{total current asset} dropped from 22M to 19M~(Figure~\ref{fig:usageScenario}g)~(\textbf{T3}).
This observation implies that the increasing of \textit{tax income ratio} and decreasing the \textit{total current assets} of NJ hurt the fiscal ranking.
Thus, the government officer understood the comparably important attributes that affected the ranking of NJ. She could ultimately focus on a few specific attributes instead of the datasets' numerous attributes for investigating the fiscal ranking of an individual state.

%% file: sections/caseStudy.tex
\begin{figure*}
\begin{center}
\includegraphics[width=0.98\textwidth]{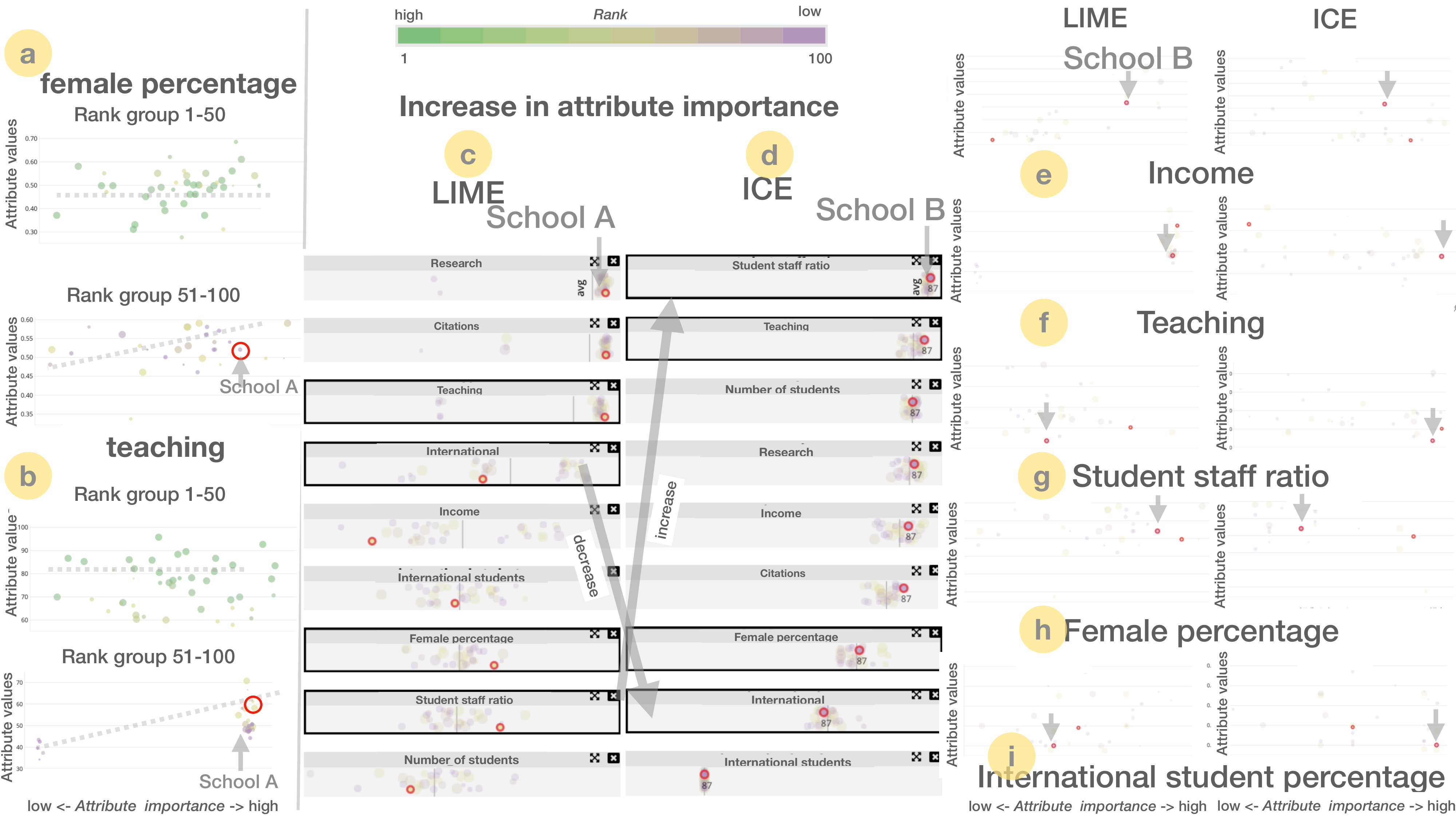}
\caption{\label{fig:caseStudy} \textbf{Usage Scenario: School rankings~(Section~\ref{sec:education}).}
a) Attribute importance correlation plots for \textit{female percentage} between the rank range 1-50 and 51-100.
b) Attribute importance correlation plots for \textit{teaching} between the rank range 1-50 and 51-100.
c) and d) The attribute importance distribution plots for \jun{rank range  51 - 100 between LIME and ICE explanations}, with highlights on schools B and A. 
e) - i) The attribute importance correlation plots for the five attributes in the \jun{rank range 51 - 100 between LIME and ICE explanations}. The x-axis is the attribute importance, and the y-axis is the attribute value.
}
\end{center}
\end{figure*}

\subsection{Making Choices for Higher Education}
\label{sec:education}

In this usage scenario, we focused on understanding how \texttt{TRIVEA}  can be used by student applicants, for whom searching for a good university is a challenging task since their priorities may not match directly with that of the universities. A good way to understand a university's priority is to understand the correlation between its yearly rankings and the factors affecting them. 
Hence, they can look for a university that matches their priority best and may be more suitable than a top-ranked university that does not suit their priorities. For this scenario, we use the university data set~\cite{University} introduced earlier in Section~\ref{sec:visualizationsdesign}.
\jun{
An applicant first examined the rankers in the range of $1$ to $100$ in the default ranker mode. Using Mean Average Precision and manually checking the rank deviations, he concluded that the Ranking SVM model performed the best. 
Then the applicant used the range comparison mode in \texttt{TRIVEA}  and chose the range $1$ - $50$ and $51$ - $100$ as the two groups for comparison using the \jun{Ranking SVM} model. Group 1 dots were green-yellow, and group 2 yellow-purple. 
He saw \textit{female percentage} was an important attribute. Using the time navigation in \texttt{TRIVEA}, he observed that it was not a high priority in both ranges over the years.
The highest priorities were \textit{research},  \textit{teaching}, and \textit{citations}
~(\textbf{T2}), which may reflect that a university emphasizes research. 
However, the basic needs for students are good education and a sense of community.
He was less interested in the most important attributes as computed by the explanations. Instead,
his priorities were \textit{female percentage}, \textit{teaching}, \textit{student staff ratio}.
He first investigated the importance of correlation plots for \textit{female percentage} and \textit{teaching}.
The \textit{female percentage} showed a no correlation in group 1~(1-50), but a positive correlation in group 2~(51-100)~(Figure~\ref{fig:caseStudy}a)~(\textbf{T3}). The higher-ranked school in group 2 mostly have high values too. Group 1 schools also have high values, but a female student may be more competitive when applying for group 2 schools because of the positive correlation.
The same correlation patterns can be observed for \textit{teaching}~(Figure~\ref{fig:caseStudy}b)~(\textbf{T3}). Although the teaching scores for group 2 were 20 percent lower than group 1 on average (shown on the y-axis of Figure~\ref{fig:caseStudy}b), there were schools that both had high teaching importance and were on par with the group 1 school teaching scores. 
According to the color gradient, \textit{teaching} correlated to the rank change in group 2~(Figure~\ref{fig:caseStudy}b). Hence, choosing a school where a high teaching score was of high importance fulfilled one of the needs of the student.
He found a school A~(Brown Uni) in group 2. Its \textit{teaching} attribute had high importance and value~(Figure~\ref{fig:caseStudy}b). 
School A appeared to be balanced in \textit{female percentage}~(Figure~\ref{fig:caseStudy}h) and  high on \textit{teaching}~(Figure~\ref{fig:caseStudy}f)~(\textbf{T2}), but all high on importance. The school matched the student's needs and was less competitive for applicants than the group 1 schools. Yet, it could compete with group 1 schools on the attributes that the student cared about. The school might also intend to improve on the same attributes since they appeared important for its ranking. But for the group 1 schools, they might devote most of their effort to \textit{research}. 
Exploring and choosing a school based on the explanation broadened the applicants' perspective on school selection. He wanted to see if using another explainer may yield additional choices. He switched the explainer from LIME to ICE.
Immediately, he noticed that ICE explanation for ranking SVM prioritizes the \textit{student staff ratio}~(\textbf{T2}), which contradicted LIME~(Figure~\ref{fig:caseStudy}c,d). 
He noticed that school A still showed a high impact in \textit{student staff ratio} in ICE. He also noticed a nearby dot, with lower \textit{student staff ratio} compared to school A, representing school B~(Vanderbilt Uni).
He chose to carefully examine if school B is better than A.
Across all absolute attribute values, school B was better than school A on \textit{student staff ratio}, \textit{female percentage}, but lower in \textit{teaching}. LIME and ICE explainers both agreed that \textit{student staff ratio} and \textit{female percentage} was more important to school A but did not agree on \textit{teaching}.
He concluded that he favors school A since the explainers' agreement favors school A over B.
Thus, he observed the difference between the top 50, 51-100
from 2011 to 2016 and found schools that matched the student's interests. As a result, he understood the advantages and disadvantages of applying to those universities in the coming year and strengthened his higher education strategy.
Leveraging different explainers, he was able to quickly identify the schools according to his interest and examine their attributes for school decisions.
}

%% file: sections/subjectivefeedback.tex
\section{Subjective Feedback from Domain Experts}

For evaluating \texttt{TRIVEA}, we asked four researchers from diverse domains, such as energy, cyber security, data science, etc., who are familiar with and use ranking applications, for their subjective feedback~\cite{lam2012empirical}.
We used questionnaires and online feedback~(using emails and discussions) to understand how people can benefit from using \texttt{TRIVEA} for interpreting the algorithmic rankers used as part of our surrogate model for explaining inaccessible ranking schemes. All four experts have doctoral degrees in computer science or engineering, with experience ranging from $5$ years to $10$ years. Two of them have experience developing or using machine learning models and are somewhat familiar with visualization tools.

We provided them with a training video where we described all the functionalities of \texttt{TRIVEA} and asked them to use it for free-form exploration of learned rankings using any of the data sets we used. We assessed their perceived ease of use, comfort, level of confidence, challenges in using \texttt{TRIVEA}, and potential shortcomings from their written and verbal feedback.

We found that all participants were comfortable using \texttt{TRIVEA} and particularly appreciated the ability to probe the model outcomes by linking uncertainty information with the explanation: \textit{``The textured bars and the dot sizes are very helpful for me to quickly filter out inaccurate ranks"}. Two of them could immediately relate to how \texttt{TRIVEA} can be used for problem-solving in cases where there is a need to learn rankings from data and explain them to augment end user's trust: \textit{``This could be helpful in the energy domain. Some use cases could be: Ranking of energy efficient different buildings"}, and \textit{``utility ranking and priority of loads are some of them come to mind”}, and \textit{``can be valuable for operators who need to balance different criteria before making decisions for operating the grid"}.

All of them noted the benefit of the flexibility the interactions like filtering and animation afforded. Two participants, who develop machine learning models as part of their research, noted how \texttt{TRIVEA} can help them in model selection: \textit{``This is a very helpful tool for ML researchers who are often confused between which ML algorithms to use for a particular task. It provides a nice visual analysis.”}

One of them mentioned the need to potentially incorporate multiple explanation techniques for an even detailed comparison of attribute contributions: \textit{``The attribute importance is well presented. The designer might consider adding more criteria for attribute importance ranking”}. Another participant noted that while the explanation plots and color-coded rankings are helpful in building a mental model of attribute contributions quickly, one might augment this view with the ability to save one's results in the interface. This comment encourages us to pursue directions such as knowledge externalization based on inferences from ensemble algorithmic rankers.

\section{Discussion}

In this section, we discuss the effectiveness of \texttt{TRIVEA} in communicating outcomes from ensemble algorithmic rankers by reflecting on the subjective feedback from participants and based on our assessment of state of the art. 
\texttt{TRIVEA} is able to encourage multi-model comparison of model fitness and explanations for evaluating and interpreting rankings. 
However, there is a performance trade-off owing to the data range and the number of models, especially when we are simultaneously analyzing rankings and explanations. 
We noted that for optimal user experience, one either selects a limited number of rankers (about $5$) or limits the data range to about $100$ when analyzing both rankings and explanations. We will address this issue in the future. 
For the animations, we noticed that augmenting more visual indicators of what is changing and the \textit{before} and \textit{after} states will be helpful in further communicating salient changes. 
On the machine learning side, we can afford to link \texttt{TRIVEA} more explicitly to model training and selection. 
While we are not re-training the models in our case, insights from \texttt{TRIVEA} can be used for such purposes and to better align a domain expert’s mental model of how an attribute contributes to rankings. 
\texttt{TRIVEA} can also provide insight into model stability across years and encourage looking at developing new metrics for calibrating performance in local neighborhoods. 
When we used the advanced learning-to-rank models~(LTR) for webpage ranking, we used simpler models, like linear regression or Random Forest, for performance comparison. 
In some cases, the latter outperformed advanced LTR models. This can be explained by the fact that advanced LTR models are data-hungry, and the size of our training data, in some cases, did not meet those requirements. However, learning to predict rankings from smaller data sets is a viable task, and \texttt{TRIVEA} provides a way to incorporate those ranking-driven inference scenarios. 
The explanation plots are now limited with respect to the quality of LIME output. However, our abstraction of LIME output can be generalized to other XAI methods for understanding dynamic local attribute importance about ranking output.  
We plan to expand the explanation methods to other XAI methods like SHAP~\cite{lundberg2017unified} and Anchor~\cite{ribeiro2018anchors}, for aiding in more generalizable inferences from observation of local attribute importance. 
Also, our interface can be adapted to explain other models, such as classifiers, by redesigning the deviation plot module in the interface.
Our work is related to the accessibility and transparency issue of the data and ranker to humans, especially the end users of the rankings.
We will continue to focus on the accessibility issue of data and rankers where data and formulas may not be completely accessible in the public domain, as is common in many socio-technical applications of ranking, like, hiring, admissions, etc.

\section{Conclusion}

In this paper, we demonstrate how the \texttt{TRIVEA} visual analytic system can aid in the interpretation of algorithmic rankers and drive user inferences for both rankees and decision-makers. This is an important contribution for making sure proprietary ranking schemes are made broadly accessible and auditable in the future. 
We enable multi-model comparisons of learned rankings and their explanations for generating user inferences. While these tasks are of high reasoning complexity, we demonstrated through the usage scenarios and expert feedback that our visualization and interface design choices, combined with filtering and animation strategies, can provide appropriate affordances for solving these tasks. We plan to conduct controlled user studies in the future further to evaluate the effectiveness of \texttt{TRIVEA}.

%% file: main_sn-article.bbl

\begin{thebibliography}{43}
\ifx \bisbn   \undefined \def \bisbn  #1{ISBN #1}\fi
\ifx \binits  \undefined \def \binits#1{#1}\fi
\ifx \bauthor  \undefined \def \bauthor#1{#1}\fi
\ifx \batitle  \undefined \def \batitle#1{#1}\fi
\ifx \bjtitle  \undefined \def \bjtitle#1{#1}\fi
\ifx \bvolume  \undefined \def \bvolume#1{\textbf{#1}}\fi
\ifx \byear  \undefined \def \byear#1{#1}\fi
\ifx \bissue  \undefined \def \bissue#1{#1}\fi
\ifx \bfpage  \undefined \def \bfpage#1{#1}\fi
\ifx \blpage  \undefined \def \blpage #1{#1}\fi
\ifx \burl  \undefined \def \burl#1{\textsf{#1}}\fi
\ifx \doiurl  \undefined \def \doiurl#1{\url{https://doi.org/#1}}\fi
\ifx \betal  \undefined \def \betal{\textit{et al.}}\fi
\ifx \binstitute  \undefined \def \binstitute#1{#1}\fi
\ifx \binstitutionaled  \undefined \def \binstitutionaled#1{#1}\fi
\ifx \bctitle  \undefined \def \bctitle#1{#1}\fi
\ifx \beditor  \undefined \def \beditor#1{#1}\fi
\ifx \bpublisher  \undefined \def \bpublisher#1{#1}\fi
\ifx \bbtitle  \undefined \def \bbtitle#1{#1}\fi
\ifx \bedition  \undefined \def \bedition#1{#1}\fi
\ifx \bseriesno  \undefined \def \bseriesno#1{#1}\fi
\ifx \blocation  \undefined \def \blocation#1{#1}\fi
\ifx \bsertitle  \undefined \def \bsertitle#1{#1}\fi
\ifx \bsnm \undefined \def \bsnm#1{#1}\fi
\ifx \bsuffix \undefined \def \bsuffix#1{#1}\fi
\ifx \bparticle \undefined \def \bparticle#1{#1}\fi
\ifx \barticle \undefined \def \barticle#1{#1}\fi
\bibcommenthead
\ifx \bconfdate \undefined \def \bconfdate #1{#1}\fi
\ifx \botherref \undefined \def \botherref #1{#1}\fi
\ifx \url \undefined \def \url#1{\textsf{#1}}\fi
\ifx \bchapter \undefined \def \bchapter#1{#1}\fi
\ifx \bbook \undefined \def \bbook#1{#1}\fi
\ifx \bcomment \undefined \def \bcomment#1{#1}\fi
\ifx \oauthor \undefined \def \oauthor#1{#1}\fi
\ifx \citeauthoryear \undefined \def \citeauthoryear#1{#1}\fi
\ifx \endbibitem  \undefined \def \endbibitem {}\fi
\ifx \bconflocation  \undefined \def \bconflocation#1{#1}\fi
\ifx \arxivurl  \undefined \def \arxivurl#1{\textsf{#1}}\fi
\csname PreBibitemsHook\endcsname

\bibitem{bauer2009designing}
\begin{bchapter}
\bauthor{\bsnm{Bauer}, \binits{J.M.}},
\bauthor{\bsnm{Herder}, \binits{P.M.}}:
\bctitle{Designing socio-technical systems}.
In: \beditor{\bsnm{Meijers}, \binits{A.}} (ed.)
\bbtitle{Philosophy of Technology and Engineering Sciences}.
\bsertitle{Handbook of the Philosophy of Science},
pp. \bfpage{601}--\blpage{630}.
\bpublisher{North-Holland},
\blocation{Amsterdam}
(\byear{2009})
\end{bchapter}
\endbibitem

\bibitem{ribeiro2016modelagnostic}
\begin{botherref}
\oauthor{\bsnm{Ribeiro}, \binits{M.T.}},
\oauthor{\bsnm{Singh}, \binits{S.}},
\oauthor{\bsnm{Guestrin}, \binits{C.}}:
Model-agnostic interpretability of machine learning.
arXiv preprint arXiv:1606.05386
(2016)
\end{botherref}
\endbibitem

\bibitem{mythos}
\begin{barticle}
\bauthor{\bsnm{Lipton}, \binits{Z.C.}}:
\batitle{The mythos of model interpretability: In machine learning, the concept
  of interpretability is both important and slippery.}
\bjtitle{Queue}
\bvolume{16}(\bissue{3}),
\bfpage{31}--\blpage{57}
(\byear{2018})
\end{barticle}
\endbibitem

\bibitem{heer2019agency}
\begin{barticle}
\bauthor{\bsnm{Heer}, \binits{J.}}:
\batitle{Agency plus automation: Designing artificial intelligence into
  interactive systems}.
\bjtitle{Proceedings of the National Academy of Sciences}
\bvolume{116}(\bissue{6}),
\bfpage{1844}--\blpage{1850}
(\byear{2019})
\end{barticle}
\endbibitem

\bibitem{shneiderman2022human}
\begin{bbook}
\bauthor{\bsnm{Shneiderman}, \binits{B.}}:
\bbtitle{Human-centered AI}.
\bpublisher{Oxford University Press},
\blocation{Oxford, UK}
(\byear{2022})
\end{bbook}
\endbibitem

\bibitem{seo2005rank}
\begin{barticle}
\bauthor{\bsnm{Seo}, \binits{J.}},
\bauthor{\bsnm{Shneiderman}, \binits{B.}}:
\batitle{A rank-by-feature framework for interactive exploration of
  multidimensional data}.
\bjtitle{Information Visualization}
\bvolume{4}(\bissue{2}),
\bfpage{96}--\blpage{113}
(\byear{2005})
\end{barticle}
\endbibitem

\bibitem{seo2006knowledge}
\begin{barticle}
\bauthor{\bsnm{Seo}, \binits{J.}},
\bauthor{\bsnm{Shneiderman}, \binits{B.}}:
\batitle{Knowledge discovery in high-dimensional data: Case studies and a user
  survey for the rank-by-feature framework}.
\bjtitle{IEEE transactions on visualization and computer graphics}
\bvolume{12}(\bissue{3}),
\bfpage{311}--\blpage{322}
(\byear{2006})
\end{barticle}
\endbibitem

\bibitem{perin2014table}
\begin{bchapter}
\bauthor{\bsnm{Perin}, \binits{C.}},
\bauthor{\bsnm{Vuillemot}, \binits{R.}},
\bauthor{\bsnm{Fekete}, \binits{J.-D.}}:
\bctitle{{\`A} table! improving temporal navigation in soccer ranking tables}.
In: \bbtitle{Proceedings of the SIGCHI Conference on Human Factors in Computing
  Systems},
pp. \bfpage{887}--\blpage{896}
(\byear{2014})
\end{bchapter}
\endbibitem

\bibitem{gratzl2013lineup}
\begin{barticle}
\bauthor{\bsnm{Gratzl}, \binits{S.}},
\bauthor{\bsnm{Lex}, \binits{A.}},
\bauthor{\bsnm{Gehlenborg}, \binits{N.}},
\bauthor{\bsnm{Pfister}, \binits{H.}},
\bauthor{\bsnm{Streit}, \binits{M.}}:
\batitle{Lineup: Visual analysis of multi-attribute rankings}.
\bjtitle{IEEE transactions on visualization and computer graphics}
\bvolume{19}(\bissue{12}),
\bfpage{2277}--\blpage{2286}
(\byear{2013})
\end{barticle}
\endbibitem

\bibitem{shi2012rankexplorer}
\begin{barticle}
\bauthor{\bsnm{Shi}, \binits{C.}},
\bauthor{\bsnm{Cui}, \binits{W.}},
\bauthor{\bsnm{Liu}, \binits{S.}},
\bauthor{\bsnm{Xu}, \binits{P.}},
\bauthor{\bsnm{Chen}, \binits{W.}},
\bauthor{\bsnm{Qu}, \binits{H.}}:
\batitle{Rankexplorer: Visualization of ranking changes in large time series
  data}.
\bjtitle{IEEE Transactions on Visualization and Computer Graphics}
\bvolume{18}(\bissue{12}),
\bfpage{2669}--\blpage{2678}
(\byear{2012})
\end{barticle}
\endbibitem

\bibitem{havre2000themeriver}
\begin{bchapter}
\bauthor{\bsnm{Havre}, \binits{S.}},
\bauthor{\bsnm{Hetzler}, \binits{B.}},
\bauthor{\bsnm{Nowell}, \binits{L.}}:
\bctitle{Themeriver: Visualizing theme changes over time}.
In: \bbtitle{IEEE Symposium on Information Visualization 2000. INFOVIS 2000.
  Proceedings},
pp. \bfpage{115}--\blpage{123}
(\byear{2000}).
\bcomment{IEEE}
\end{bchapter}
\endbibitem

\bibitem{miranda2017topkube}
\begin{barticle}
\bauthor{\bsnm{Miranda}, \binits{F.}},
\bauthor{\bsnm{Lins}, \binits{L.}},
\bauthor{\bsnm{Klosowski}, \binits{J.T.}},
\bauthor{\bsnm{Silva}, \binits{C.T.}}:
\batitle{Topkube: A rank-aware data cube for real-time exploration of
  spatiotemporal data}.
\bjtitle{IEEE Transactions on visualization and computer graphics}
\bvolume{24}(\bissue{3}),
\bfpage{1394}--\blpage{1407}
(\byear{2017})
\end{barticle}
\endbibitem

\bibitem{wall2017podium}
\begin{barticle}
\bauthor{\bsnm{Wall}, \binits{E.}},
\bauthor{\bsnm{Das}, \binits{S.}},
\bauthor{\bsnm{Chawla}, \binits{R.}},
\bauthor{\bsnm{Kalidindi}, \binits{B.}},
\bauthor{\bsnm{Brown}, \binits{E.T.}},
\bauthor{\bsnm{Endert}, \binits{A.}}:
\batitle{Podium: Ranking data using mixed-initiative visual analytics}.
\bjtitle{IEEE transactions on visualization and computer graphics}
\bvolume{24}(\bissue{1}),
\bfpage{288}--\blpage{297}
(\byear{2017})
\end{barticle}
\endbibitem

\bibitem{joachims2002optimizing}
\begin{bchapter}
\bauthor{\bsnm{Joachims}, \binits{T.}}:
\bctitle{Optimizing search engines using clickthrough data}.
In: \bbtitle{Proceedings of the Eighth ACM SIGKDD International Conference on
  Knowledge Discovery and Data Mining},
pp. \bfpage{133}--\blpage{142}
(\byear{2002})
\end{bchapter}
\endbibitem

\bibitem{ma2019explaining}
\begin{barticle}
\bauthor{\bsnm{Ma}, \binits{Y.}},
\bauthor{\bsnm{Xie}, \binits{T.}},
\bauthor{\bsnm{Li}, \binits{J.}},
\bauthor{\bsnm{Maciejewski}, \binits{R.}}:
\batitle{Explaining vulnerabilities to adversarial machine learning through
  visual analytics}.
\bjtitle{IEEE transactions on visualization and computer graphics}
\bvolume{26}(\bissue{1}),
\bfpage{1075}--\blpage{1085}
(\byear{2019})
\end{barticle}
\endbibitem

\bibitem{krause2017workflow}
\begin{bchapter}
\bauthor{\bsnm{Krause}, \binits{J.}},
\bauthor{\bsnm{Dasgupta}, \binits{A.}},
\bauthor{\bsnm{Swartz}, \binits{J.}},
\bauthor{\bsnm{Aphinyanaphongs}, \binits{Y.}},
\bauthor{\bsnm{Bertini}, \binits{E.}}:
\bctitle{A workflow for visual diagnostics of binary classifiers using
  instance-level explanations}.
In: \bbtitle{IEEE Conference on Visual Analytics Science and Technology
  (VAST)},
pp. \bfpage{162}--\blpage{172}
(\byear{2017}).
\bcomment{IEEE}
\end{bchapter}
\endbibitem

\bibitem{gomez2021advice}
\begin{bchapter}
\bauthor{\bsnm{Gomez}, \binits{O.}},
\bauthor{\bsnm{Holter}, \binits{S.}},
\bauthor{\bsnm{Yuan}, \binits{J.}},
\bauthor{\bsnm{Bertini}, \binits{E.}}:
\bctitle{Advice: Aggregated visual counterfactual explanations for machine
  learning model validation}.
In: \bbtitle{IEEE Visualization Conference (VIS)},
pp. \bfpage{31}--\blpage{35}
(\byear{2021}).
\bcomment{IEEE}
\end{bchapter}
\endbibitem

\bibitem{chan2020subplex}
\begin{barticle}
\bauthor{\bsnm{Yuan}, \binits{J.}},
\bauthor{\bsnm{Chan}, \binits{G.Y.-Y.}},
\bauthor{\bsnm{Barr}, \binits{B.}},
\bauthor{\bsnm{Overton}, \binits{K.}},
\bauthor{\bsnm{Rees}, \binits{K.}},
\bauthor{\bsnm{Nonato}, \binits{L.G.}},
\bauthor{\bsnm{Bertini}, \binits{E.}},
\bauthor{\bsnm{Silva}, \binits{C.T.}}:
\batitle{Subplex: A visual analytics approach to understand local model
  explanations at the subpopulation level}.
\bjtitle{IEEE Computer Graphics and Applications}
\bvolume{42}(\bissue{6}),
\bfpage{24}--\blpage{36}
(\byear{2022})
\end{barticle}
\endbibitem

\bibitem{xu2022visual}
\begin{botherref}
\oauthor{\bsnm{Xu}, \binits{X.}},
\oauthor{\bsnm{Mo}, \binits{J.}}:
Visual explanation and robustness assessment optimization of saliency maps for
  image classification.
The Visual Computer,
1--17
(2022)
\end{botherref}
\endbibitem

\bibitem{gleicher2020boxer}
\begin{bchapter}
\bauthor{\bsnm{Gleicher}, \binits{M.}},
\bauthor{\bsnm{Barve}, \binits{A.}},
\bauthor{\bsnm{Yu}, \binits{X.}},
\bauthor{\bsnm{Heimerl}, \binits{F.}}:
\bctitle{Boxer: Interactive comparison of classifier results}.
In: \bbtitle{Computer Graphics Forum},
vol. \bseriesno{39},
pp. \bfpage{181}--\blpage{193}
(\byear{2020}).
\bcomment{Wiley Online Library}
\end{bchapter}
\endbibitem

\bibitem{zhang2018manifold}
\begin{barticle}
\bauthor{\bsnm{Zhang}, \binits{J.}},
\bauthor{\bsnm{Wang}, \binits{Y.}},
\bauthor{\bsnm{Molino}, \binits{P.}},
\bauthor{\bsnm{Li}, \binits{L.}},
\bauthor{\bsnm{Ebert}, \binits{D.S.}}:
\batitle{Manifold: A model-agnostic framework for interpretation and diagnosis
  of machine learning models}.
\bjtitle{IEEE transactions on visualization and computer graphics}
\bvolume{25}(\bissue{1}),
\bfpage{364}--\blpage{373}
(\byear{2018})
\end{barticle}
\endbibitem

\bibitem{arendt2020crosscheck}
\begin{botherref}
\oauthor{\bsnm{Arendt}, \binits{D.}},
\oauthor{\bsnm{Huang}, \binits{Z.}},
\oauthor{\bsnm{Shrestha}, \binits{P.}},
\oauthor{\bsnm{Ayton}, \binits{E.}},
\oauthor{\bsnm{Glenski}, \binits{M.}},
\oauthor{\bsnm{Volkova}, \binits{S.}}:
Crosscheck: Rapid, reproducible, and interpretable model evaluation.
arXiv preprint arXiv:2004.07993
(2020)
\end{botherref}
\endbibitem

\bibitem{ribeiro2016should}
\begin{bchapter}
\bauthor{\bsnm{Ribeiro}, \binits{M.T.}},
\bauthor{\bsnm{Singh}, \binits{S.}},
\bauthor{\bsnm{Guestrin}, \binits{C.}}:
\bctitle{" why should i trust you?" explaining the predictions of any
  classifier}.
In: \bbtitle{Proceedings of the 22nd ACM SIGKDD International Conference on
  Knowledge Discovery and Data Mining},
pp. \bfpage{1135}--\blpage{1144}
(\byear{2016})
\end{bchapter}
\endbibitem

\bibitem{hong2020human}
\begin{barticle}
\bauthor{\bsnm{Hong}, \binits{S.R.}},
\bauthor{\bsnm{Hullman}, \binits{J.}},
\bauthor{\bsnm{Bertini}, \binits{E.}}:
\batitle{Human factors in model interpretability: Industry practices,
  challenges, and needs}.
\bjtitle{Proceedings of the ACM on Human-Computer Interaction}
\bvolume{4}(\bissue{CSCW1}),
\bfpage{1}--\blpage{26}
(\byear{2020})
\end{barticle}
\endbibitem

\bibitem{gleicher2017considerations}
\begin{barticle}
\bauthor{\bsnm{Gleicher}, \binits{M.}}:
\batitle{Considerations for visualizing comparison}.
\bjtitle{IEEE transactions on visualization and computer graphics}
\bvolume{24}(\bissue{1}),
\bfpage{413}--\blpage{423}
(\byear{2017})
\end{barticle}
\endbibitem

\bibitem{dasgupta2019separating}
\begin{barticle}
\bauthor{\bsnm{Dasgupta}, \binits{A.}},
\bauthor{\bsnm{Wang}, \binits{H.}},
\bauthor{\bsnm{O'Brien}, \binits{N.}},
\bauthor{\bsnm{Burrows}, \binits{S.}}:
\batitle{Separating the wheat from the chaff: Comparative visual cues for
  transparent diagnostics of competing models}.
\bjtitle{IEEE transactions on visualization and computer graphics}
\bvolume{26}(\bissue{1}),
\bfpage{1043}--\blpage{1053}
(\byear{2020})
\end{barticle}
\endbibitem

\bibitem{yuan2022rankers}
\begin{botherref}
\oauthor{\bsnm{Yuan}, \binits{J.}},
\oauthor{\bsnm{Stoyanovich}, \binits{J.}},
\oauthor{\bsnm{Dasgupta}, \binits{A.}}:
Rankers, rankees, \& rankings: Peeking into the pandora's box from a
  socio-technical perspective.
CHI Workshop on Human-Centered Data Science~(HCDS),
2211
(2022)
\end{botherref}
\endbibitem

\bibitem{page1999pagerank}
\begin{botherref}
\oauthor{\bsnm{Page}, \binits{L.}},
\oauthor{\bsnm{Brin}, \binits{S.}},
\oauthor{\bsnm{Motwani}, \binits{R.}},
\oauthor{\bsnm{Winograd}, \binits{T.}}:
The pagerank citation ranking: Bringing order to the web.
Technical report,
Stanford InfoLab
(1999)
\end{botherref}
\endbibitem

\bibitem{University}
\begin{botherref}
Times World University Rankings.
\url{https://www.kaggle.com/mylesoneill/world-university-rankings#timesData.csv}.
Accessed: 2020-04-30
\end{botherref}
\endbibitem

\bibitem{liu2015application}
\begin{barticle}
\bauthor{\bsnm{Liu}, \binits{B.}},
\bauthor{\bsnm{Chen}, \binits{J.}},
\bauthor{\bsnm{Wang}, \binits{X.}}:
\batitle{Application of learning to rank to protein remote homology detection}.
\bjtitle{Bioinformatics}
\bvolume{31}(\bissue{21}),
\bfpage{3492}--\blpage{3498}
(\byear{2015})
\end{barticle}
\endbibitem

\bibitem{mohler2020learning}
\begin{barticle}
\bauthor{\bsnm{Mohler}, \binits{G.}},
\bauthor{\bsnm{Porter}, \binits{M.}},
\bauthor{\bsnm{Carter}, \binits{J.}},
\bauthor{\bsnm{LaFree}, \binits{G.}}:
\batitle{Learning to rank spatio-temporal event hotspots}.
\bjtitle{Crime Science}
\bvolume{9}(\bissue{1}),
\bfpage{1}--\blpage{12}
(\byear{2020})
\end{barticle}
\endbibitem

\bibitem{lundberg2017unified}
\begin{botherref}
\oauthor{\bsnm{Lundberg}, \binits{S.M.}},
\oauthor{\bsnm{Lee}, \binits{S.-I.}}:
A unified approach to interpreting model predictions.
Advances in neural information processing systems
\textbf{30}
(2017)
\end{botherref}
\endbibitem

\bibitem{yeh2022bringing}
\begin{bchapter}
\bauthor{\bsnm{Yeh}, \binits{A.}},
\bauthor{\bsnm{Ngo}, \binits{A.}}:
\bctitle{Bringing a ruler into the black box: Uncovering feature impact from
  individual conditional expectation plots}.
In: \bbtitle{Machine Learning and Principles and Practice of Knowledge
  Discovery in Databases: International Workshops of ECML PKDD 2021, Virtual
  Event, September 13-17, 2021, Proceedings, Part I},
pp. \bfpage{34}--\blpage{48}
(\byear{2022}).
\bcomment{Springer}
\end{bchapter}
\endbibitem

\bibitem{RankLib}
\begin{botherref}
RankLib Toolkit.
\url{https://sourceforge.net/p/lemur/wiki/RankLib/}.
Accessed: 2020-04-29
\end{botherref}
\endbibitem

\bibitem{LimeRepo}
\begin{botherref}
LIME Python Package.
\url{https://github.com/marcotcr/lime}.
Accessed: 2020-04-29
\end{botherref}
\endbibitem

\bibitem{IceRepo}
\begin{botherref}
ICE feature impact Python Package.
\url{https://github.com/mixerupper/ice_feature_impact}.
Accessed: 2020-04-29
\end{botherref}
\endbibitem

\bibitem{Fiscal}
\begin{botherref}
State Fiscal Rankings.
\url{https://www.mercatus.org/system/files/masterfiscalrankingsdata2006-2016.xlsx}.
Accessed: 2020-04-30
\end{botherref}
\endbibitem

\bibitem{pang1997approaches}
\begin{barticle}
\bauthor{\bsnm{Pang}, \binits{A.T.}},
\bauthor{\bsnm{Wittenbrink}, \binits{C.M.}},
\bauthor{\bsnm{Lodha}, \binits{S.K.}}, \betal:
\batitle{Approaches to uncertainty visualization}.
\bjtitle{The Visual Computer}
\bvolume{13}(\bissue{8}),
\bfpage{370}--\blpage{390}
(\byear{1997})
\end{barticle}
\endbibitem

\bibitem{maack2022uncertainty}
\begin{botherref}
\oauthor{\bsnm{Maack}, \binits{R.G.}},
\oauthor{\bsnm{Scheuermann}, \binits{G.}},
\oauthor{\bsnm{Hagen}, \binits{H.}},
\oauthor{\bsnm{Pe{\~n}aloza}, \binits{J.T.H.}},
\oauthor{\bsnm{Gillmann}, \binits{C.}}:
Uncertainty-aware visual analytics: scope, opportunities, and challenges.
The Visual Computer,
1--22
(2022)
\end{botherref}
\endbibitem

\bibitem{valizadegan2009learning}
\begin{bchapter}
\bauthor{\bsnm{Valizadegan}, \binits{H.}},
\bauthor{\bsnm{Jin}, \binits{R.}},
\bauthor{\bsnm{Zhang}, \binits{R.}},
\bauthor{\bsnm{Mao}, \binits{J.}}:
\bctitle{Learning to rank by optimizing ndcg measure.}
In: \bbtitle{NIPS},
vol. \bseriesno{22},
pp. \bfpage{1883}--\blpage{1891}
(\byear{2009})
\end{bchapter}
\endbibitem

\bibitem{robertson2008new}
\begin{bchapter}
\bauthor{\bsnm{Robertson}, \binits{S.}}:
\bctitle{A new interpretation of average precision}.
In: \bbtitle{Proceedings of the 31st Annual International ACM SIGIR Conference
  on Research and Development in Information Retrieval},
pp. \bfpage{689}--\blpage{690}
(\byear{2008})
\end{bchapter}
\endbibitem

\bibitem{lam2012empirical}
\begin{barticle}
\bauthor{\bsnm{Lam}, \binits{H.}},
\bauthor{\bsnm{Bertini}, \binits{E.}},
\bauthor{\bsnm{Isenberg}, \binits{P.}},
\bauthor{\bsnm{Plaisant}, \binits{C.}},
\bauthor{\bsnm{Carpendale}, \binits{S.}}:
\batitle{Empirical studies in information visualization: Seven scenarios}.
\bjtitle{IEEE transactions on visualization and computer graphics}
\bvolume{18}(\bissue{9}),
\bfpage{1520}--\blpage{1536}
(\byear{2011})
\end{barticle}
\endbibitem

\bibitem{ribeiro2018anchors}
\begin{bchapter}
\bauthor{\bsnm{Ribeiro}, \binits{M.T.}},
\bauthor{\bsnm{Singh}, \binits{S.}},
\bauthor{\bsnm{Guestrin}, \binits{C.}}:
\bctitle{Anchors: High-precision model-agnostic explanations}.
In: \bbtitle{Proceedings of the AAAI Conference on Artificial Intelligence},
vol. \bseriesno{32}
(\byear{2018})
\end{bchapter}
\endbibitem

\end{thebibliography}
